\documentclass[english,aps,prb,twocolumn,amsmath,showpacs,superscriptaddress,preprintnumbers]{revtex4}

\usepackage{amsmath}
\usepackage{amssymb}
\usepackage{mathrsfs}
\usepackage{graphicx}
\usepackage{enumerate}
\usepackage{epstopdf}
\usepackage{slashed}
\usepackage{dsfont}

\def\e{\epsilon}

\def\d{\partial}
\def\rg{\mathrm{g}}

\def\sl2r{SL(2,\mathbb{R})}
\DeclareMathOperator{\Tr}{Tr}

\def\sl2r{SL(2,\mathbb{R})}
\def\zb{{\bar z}}

\def\Tr{\mbox{Tr}\,}
\newcommand{\la}{\label}

\newcommand{\bea}{\begin{eqnarray}}
\newcommand{\eea}{\end{eqnarray}}
\newcommand{\p}{\partial}

\newcommand{\lsim}{\mathrel{\hbox{\rlap{\lower.55ex \hbox{$\sim$}} \kern-.3em \raise.4ex \hbox{$<$}}}}
\newcommand{\gsim}{\mathrel{\hbox{\rlap{\lower.55ex \hbox{$\sim$}} \kern-.3em \raise.4ex \hbox{$>$}}}}

\newcommand{\be}{\begin{equation}}
\newcommand{\ee}{\end{equation}}
\renewcommand\[{\begin{equation}}
\renewcommand\]{\end{equation}}

\newcommand{\ket}[1]{\left| #1 \right>} 
\newcommand{\bra}[1]{\left< #1 \right|} 

\newcommand{\sign}{\mathrm{sgn}} 

\usepackage{bm}

\def\eq#1 { \begin{equation} #1 \end{equation} }

\def\e{\epsilon}

\def\W{\Omega}

\usepackage{color}

\begin{document}
	
\title{\begin{flushright}\vspace{-1in}
			\mbox{\normalsize  EFI-16-21}
		\end{flushright}
	Exact Electromagnetic Response of Landau Level Electrons \vskip 20pt
	 }

	\date{\today}
	
	\author{Dung Xuan Nguyen}
	\email{nxdung86@uchicago.edu}
	\affiliation{Department of Physics, University of Chicago, Chicago, Illinois 60637, USA}
	\author{Andrey Gromov}
	\email{gromovand@uchicago.edu}
	
    \affiliation{Kadanoff Center for Theoretical Physics, University of Chicago, Chicago, Illinois 60637, USA}
		\begin{abstract}
 We present a simple method that allows to calculate the electromagnetic response of non-interacting electrons in strong magnetic field to arbitrary order in the gradients of external electric and magnetic fields. We illustrate the method on both non-relativistic and massless Dirac electrons filling $N$ Landau levels. First, we derive an exact relation between the electromagnetic response of the non-relativistic and Dirac electrons in the lowest Landau level. Next, we obtain a closed form expression for the polarization operator in the large $N$ (or weak magnetic field) limit. We explicitly show that in the large $N$ limit the random phase approximation (RPA) computation of the polarization tensor agrees - in leading and sub-leading order in $N$ - with a Fermi liquid computation to {\it all} orders in the gradient expansion and for arbitrary value of the $\mathrm{g}$-factor. Finally, we show that in the large $N$ limit the non-relativistic polarization tensor agrees with Dirac's in the leading and sub-leading orders in $N$, provided that Berry phase of the Dirac cone is taken into account via replacement $N\longrightarrow N+\frac{1}{2}$.		 
		\end{abstract}
	\maketitle

	\newpage
		
\section{Introduction}

\subsection{Electrons in magnetic field}
Two-dimensional electrons subject to the strong external magnetic field organize into $N$ highly degenerate Landau levels. Such many-body states are gapped when the chemical potential lies anywhere between the Landau levels and exhibit the same qualitative behavior when electrons have either Dirac or non-relativistic nature. While qualitative features such as quantized Hall conductance and absence of the ground state degeneracy on a torus are identical, there is a quantitative difference in the local linear response functions. Detailed investigation of these fine distinctions as well as certain universalities in the behavior of both lowest Landau level and large $N$ limit of the linear response functions is the objective of the present paper. Additionally, our results should be useful in the analysis of interacting FQH states using the composite fermion \cite{lopez1991fractional} and boson\cite{zhang1989effective} approaches.

 We will study the electromagnetic response of Landau level electrons in great detail. To start we present a straightforward method that allows to calculate linear response functions in the form of the generating functional for both relativistic and non-relativistic electrons filling an arbitrary number of Landau levels. In the non-relativistic case some of the results are available\cite{chen1989anyon, randjbar1990chern, lopez1991fractional}, however we present a simpler method of derivation as well as provide a number of new results. This method was first used in Ref.[\onlinecite{abanov2014electromagnetic}], but only a few results were presented. We will explain in detail how to calculate the linear response to arbitrary order in the expansion in momentum and frequency and give a compact expression for the polarization tensor in both non-relativistic and Dirac cases. In addition to the general expressions we present the leading order corrections in momentum and frequency expansion for all linear response functions in explicit form.
 
\begin{figure}
\centerline{\includegraphics[scale=0.38]{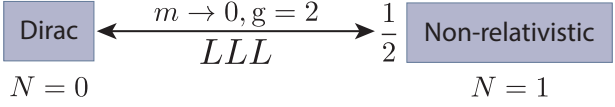}}
\caption{The linear response of the LLL is universal. The generating functionals of non-relativistic and Dirac (upon discarding divergent terms) electrons differ from each other by a factor of $\frac{1}{2}$ to {\it all} orders in the gradient expansion.  }
\label{Lowest}
\end{figure}

 With the exact expressions at hand we will investigate the linear response of the lowest Landau level (LLL). It turns out that linear response of the Dirac electrons can be extracted from the linear response of the non-relativistic electrons via simple relation \eqref{DvsNRW}, which is valid to {\it all} orders of the gradient expansion. We check this relation via an explicit computation as well as using the well-known relation between momentum-dependent Hall conductivity and the static structure factor. For the reader's convenience we summarize the universality of the LLL in Fig. \ref{Lowest}.

\begin{figure}
\centerline{\includegraphics[scale=0.44]{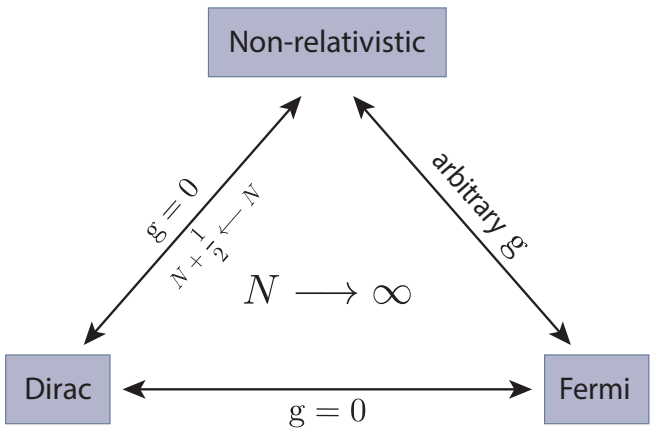}}
\caption{In the large $N$ limit electromagnetic linear response of non-relativistic electrons agrees with the response of the Fermi liquid to all orders in gradient expansion and arbitrary $\mathrm g$-factor. Electromagnetic response of Dirac electrons (in the large $N$ limit) can be extracted from either non-relativistic or Fermi liquid result upon setting $\mathrm g=0$ and replacing $N\longrightarrow N+\frac{1}{2}$. The replacement is needed to account for the contribution of the $\pi$ Berry phase of the Dirac cone. }
\label{largen}
\end{figure}

 Next, we will meticulously investigate the validity of the semiclassical approximation in the large $N$ limit. Our results on the large $N$ limit are summarized in Fig \ref{largen}. In this limit the electrons form a Fermi sphere and experience a weak magnetic field. The linear response can be calculated either using Landau's Fermi liquid (FL) theory or by directly taking the large $N$ limit of the exact expressions. We will explain how to include a finite $\mathrm{g}$-factor $\rg$ into the Fermi liquid theory and evaluate the polarization tensor exactly. We will find that in the non-relativistic case the Fermi liquid and direct large $N$ limit agree in the leading and sub-leading order in $N$ to {\it all} orders in the gradient expansion and for {\it arbitrary} value of the $\mathrm{g}$-factor (provided the latter was correctly accounted for in the FL theory, which we explain how to do). The large $N$ limit of polarization operator of Dirac electrons agrees in leading and sub-leading order in $N$ with the Fermi liquid and non-relativistic results after the Berry phase of the Dirac cone is taken into account for the value of the $\mathrm{g}$-factor $\rg=0$ (this may come as a surprise since Dirac electrons in vacuum correspond to $\rg=2$). The Fermi liquid computation is done using the novel approach of Ref. [\onlinecite{GNRS1}] where the Boltzmann equation is phrased in terms of the (bosonic) fluctuations of the shape of the Fermi surface. This formulation allows to effortlessly obtain the large $N$ polarization tensor to {\it all} orders in momentum and frequency in a closed form. We also explain how to add the effects of the short range interactions.

\subsection{Generalities}
Now we will introduce the main objects of interest, mainly to fix the notations. Given an action $S[\psi,\psi^\dag;A_\mu]$ describing the (relativistic or non-relativistic, bosonic or fermionic) charged matter fields $\psi$, coupled to external electromagnetic field $A_\mu= \bar A_\mu + \delta A_\mu$, we define the generating functional as follows
\be
W[\delta A_\mu] = - i \ln \int \mathscr D \psi \mathscr D \psi^\dag  e^{iS[\psi, \psi^\dag;\bar A_\mu+\delta A_\mu]}\,,
\ee
where $\bar A_\mu$ is the background value of the vector potential chosen to fix the chemical potential $\bar A_0 = \mu$ and background magnetic field $\epsilon^{ij}\p_i\bar A_j = \bar B = \ell^{-2}$.

Generating functional is a compact way to encode the multipoint correlation functions via
\be
\left\langle \prod_{i=1}^n J^\mu(x_i)\right\rangle = \prod_{i=1}^n \frac{\delta}{\delta \,\delta A_\mu(x_i)} W[\delta  A_\mu]\,.
\ee
The correlation functions obtained this way are always time-ordered.

In the present paper we will be interested in the {\it linear} response functions, {\it i.e.} the two-point functions with perturbations of the external fields turned off. For example, the conductivity tensor encodes linear response of electric current to the electric field and is given by
\be
\sigma^{\mu\nu}(x_1,x_2) =\left.\left(\frac{\delta}{\delta \,\delta A_\mu(x_1)} \frac{\delta}{\delta \,\delta A_\nu(x_2)}  W[\delta A_\mu]\right)\right|_{\delta A_\mu=0}\,.
\ee	
Assuming that $S[\psi, \psi^\dag;\bar A_\mu]$ describes a physical system with a spectral gap and the perturbations $\delta A_\mu$ are weak and slowly varying on the spatial scale of magnetic length $\ell$ and time scale of cyclotron frequency $\omega_c$, we can expand the generating functional $W[\delta A_\mu]$ in powers of external fields and in the gradients of external fields $\delta A_\mu$. If we also assume translational invariance then the gradient expansion can be converted into the expansion in momentum $\mathbf k$ and frequency $\Omega$. To study the linear response functions we need to keep only the terms quadratic in $\delta A_\mu$, but to arbitrary order in momentum and frequency. The most general expansion of this form is
\bea\nonumber
W[\delta A_\mu]&& = \int \frac{d\Omega d^2 \mathbf k}{(2\pi)^3} \left[\bar \rho\delta A_0\right . \\\la{WDEF}
&&\left. + \frac{1}{2} \delta A_\mu(-\mathbf k,\Omega) \Pi^{\mu\nu}(\mathbf k, \Omega) \delta A_\nu(-\mathbf k, \Omega) \right]\,,
\eea
where the matrix $\Pi^{\mu\nu}(\mathbf k, \Omega)$ is known as the polarization operator or polarization tensor. Each entry of this $3 \times 3$ matrix is an infinite double expansion in momentum and frequency. We have also implicitly used in Eq.\eqref{WDEF} that there are no currents in the ground sate of a rotational invariant system. Gauge invariance implies a Ward Identity
\be\la{gaugeWI}
\Omega \Pi^{0\mu}(\mathbf k, \Omega) + k_i \Pi^{i\mu}(\mathbf k, \Omega)=0\,.
\ee
It is easy to see that conductivity tensor is expressed in terms of the polarization tensor as
\be\la{sigmaToPi}
\sigma^{\mu\nu}(\mathbf k, \Omega) = \frac{1}{i \Omega} \Pi^{\mu\nu}(\mathbf k, \Omega)\,.
\ee

The plan the paper is as follows. We will calculate the polarization tensor $\Pi^{\mu\nu}(\mathbf k, \Omega)$ for non-relativistic electrons filling $N$ Landau levels in Section II. The main result of the Section II is the exact expression for the polarization tensor \eqref{Polanswer}. In the Section III we will calculate the polarization tensor for massless Dirac electrons filling $N$ Landau levels and compare it with the non-relativistic one in the large $N$ limit.  In the Section IV we investigate the electromagnetic response of the lowest Landau level and find an exact relation between the linear response functions for non-relativistic and Dirac electrons. In the Section V we will obtain a closed form expression for the large $N$ polarization tensor for non-relativistic electrons using the exact result \eqref{Polanswer} and using the FL theory. We find that both approaches agree exactly and differ from Dirac electrons by the contribution of the Berry phase of the Dirac cone. For reader's convenience in every Section we present an explicit form of the polarization tensor in the leading and sub-leading orders in momentum and frequency that can be understood and used without reading the rest of the paper. Various Appendices are devoted to (often tedious) technical details.

\section{Non-relativistic electrons}
\label{sec:method}
 In this Section we will explain the method for calculation of the polarization operator for the non-relativistic electrons filling $N$ Landau levels.
 \subsection{Model}
Our starting point is the system of two-dimensional non-interacting non-relativistic fermions in external electromagnetic field described by a $U(1)$ vector potential $A_{\mu}$. The action has a form
\bea \la{NR-action}
	S_{nr} =\int d^2x dt \Big[ i\hbar \psi^{\dag}D_0\psi  
	-\frac{\hbar^2}{2m}|D\psi|^2\Big] \,.
\eea
We assume that the fermions are spin polarized and, consequently, $\psi(x,t)$ is a complex Grassmann scalar. The covariant derivative 
\be\la{CovDerivative}
D_\mu = \p_\mu - i\frac{e}{\hbar c}(\bar{A}_\mu+\delta A_\mu)
\ee
includes both background vector potential and a weak perturbation. We will omit the chemical potential from the equations, but it will be implicitly assumed that the first $N$ Landau levels are completely filled in the ground state and the chemical potential lies anywhere in the gap. The natural units $\hbar = c = e =1$ are used throughout the paper. 

\subsection{Computation of the Generating Functional}
We will compute the generating functional as a gradient expansions in the external fields. Throughout the computation we will only keep the terms quadratic in the external fields, but to arbitrary order in the gradients. This expansion is well-defined because of there is a cyclotron gap in the energy spectrum. The gradient expansion can be viewed as the expansion in the inverse gap and magnetic length $\ell ^2$ which is small compared to any other spatial scale in the problem. 

\begin{figure}
\centerline{\includegraphics[scale=0.36]{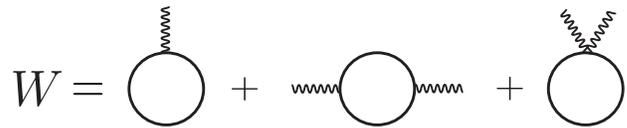}}
\caption{The generating functional to quadratic order in external fields is given by the sum of three diagrams. The first diagram, $W^{(1)}$, is linear in the perturbations of electromagnetic field and describes the constant background density of electrons. The second diagram, $W^{(2)}$, contains the main contribution to the generating functional, including the Chern-Simons term. Finally, the third diagram, $W^{(2)}_{ c}$, contains the contact terms. Note, that the last diagram vanishes for the Dirac electrons. }
\label{Diagrams}
\end{figure}

We start with rewriting the action as a differential operator sandwiched between the fermionic fields
\be
S_{nr} = \int d^2x dt \,\,\psi^\dag G^{-1} \psi\, ,
\ee
where $G^{-1}$ is the differential operator obtained by integrating by parts the derivatives acting on $\psi^\dag$. Since we assume that the perturbations of external fields are small we can write 
\be
G^{-1} = G^{-1}_0 + V\, ,
\ee
where $G^{-1}_0$ is the ``bare'' Green's function given by
\be
G^{-1}_0 = i\p_0- \frac{1}{2m} |\bar D|^2 \,,
\ee 
where $ \bar D_\mu = \p_\mu - i\bar{A}_\mu$ and 
$V$ encodes the terms at least linear in the perturbations of the external fields.
\be
V = \delta A_0 + \frac{1}{2m} |\p - i\delta A|^2\,.
\ee
Since the functional integral is quadratic in the external fields it can be formally written as a determinant of the perturbed (differential) operator $G^{-1}$. The generating functional of (connected) correlations functions is
\bea\nonumber
&&W_{\rm nr}[\delta A_\mu]= \frac{1}{i} \ln \int \mathscr D \bar\psi \mathscr D\psi \,e^{iS_{nr}[\psi, \delta A_\mu]} =\frac{1}{i} \ln \det [G^{-1}] \\
&&= -\frac{1}{i}\ln G_0 +  \frac{1}{i}\Tr(G_0V)-\frac{1}{2}\frac{1}{i}\Tr(G_0VG_0V)+\ldots\, ,
\eea
where in the last line we kept only the terms that contribute to the linear response. We can also disregard the (diverging) first term in the last line since it will not contribute to the linear response because it does not depend on the perturbations of the external fields by construction. To summarize, the object we are interested in is given by
\be
\la{W-pert}
W_{\rm nr} = W^{(1)}_{\rm nr} + W^{(2)}_{\rm nr} + W^{(2)}_{c, \rm nr} + \ldots\,,
\ee
where $W^{(1)}_{\rm nr}$ and $W^{(2)}_{\rm nr}$ are the terms linear and quadratic in external fields correspondingly, while $W_{c,\rm nr}^{(2)}$ contains the so-called contact terms (See Fig. \ref{Diagrams}).

\subsection{Fock Representation}
The Hilbert space of a particle in magnetic field can be mapped to the Hilbert space of two decoupled harmonic oscillators. To make this manifest we will use Fock representation for the basis states instead of the coordinate representation. The advantage of this approach is that we do not need to fix the gauge, thus our results will be manifestly gauge invariant. We will work in complex coordinates $z=x+iy$.

Define the creation and annihilation operators
\bea\la{adef}
&&a = \frac{i}{\sqrt{2}} \ell  \bar D_{\bar z} =\frac{i}{\sqrt{2}}\ell (\bar D_1 + i\bar D_2)\,, \\
&&a^{\dag} = \frac{i}{\sqrt{2}} \ell  \bar D_z=\frac{i}{\sqrt{2}}\ell(\bar D_1 - i\bar D_2)\,.
\eea
The inverse relations are
\be
 \bar D_z = -i\frac{1}{\sqrt{2}\ell}a^\dag\,, \quad \bar D_{\bar z} = - i\frac{1}{\sqrt{2}\ell}a
\ee
It can be easily verified that 
\be
[a,a^\dag]=1\,.
\ee 

In terms of these operators the inverse Green's function takes form
\bea\nonumber
G^{-1}_0&=&i\hbar\p_0-\omega_c\left(a^\dag a + \frac{1}{2}\right) \\
&=& i \p_0 - H_0\, ,
\eea 
where $H_0 = \omega_c\left(a^\dag a + \frac{1}{2}\right)$ is the Hamiltonian for of the particle in magnetic field.

We also define one more oscillator via
\be
b^\dag = -a+\frac{i}{\sqrt{2}\ell}z\,, \quad b = -a^\dag-\frac{i}{\sqrt{2}\ell }\bar{z}\,.\,
\ee
It can be verified that $[b,b^\dag] = 1$ and all $a$'s commute with all $b$'s. 

Operators $a^\dag, b^\dag$ generate the entire Hilbert space of the single particle problem. From this point of view the coordinates themselves must be understood as operators acting on the Hilbert space according to
\be\la{zab}
z = -\sqrt{2}\ell i(b^\dag+a)\,,\quad \bar{z} = \sqrt{2}\ell i(a^\dag+b)\,.
\ee

The basis in the Hilbert space is given by
\be\la{FockStates}
|nm\rangle = |n\rangle \otimes |m\rangle = \frac{(a^\dag)^n}{\sqrt{n!}} \frac{(b^\dag)^m}{\sqrt{m!}}|0\rangle \otimes |0\rangle\,.
\ee
The $a$-operators induce the transitions between the Landau levels, whereas $b$-operators generate the states of different angular momentum within each Landau level since
\be
[H_0, b] = [H_0, b^\dag] = 0 \,.
\ee

The bare Green's function is then given by
\be
G_0 =\int \frac{d\Omega}{2\pi} \sum_{nm}e^{-i\Omega t} \frac{|nm\rangle\langle nm|}{\Omega-E_n}\, ,
\ee
where 
\be
E_n = \left(n+\frac{1}{2}\right)\omega_c
\ee
 is the spectrum of the unperturbed Hamiltonian $H_0$.

It is easy to check that 
\be
G_0^{-1} G_0= \Big[i \p_0 - H_0 \Big] G_0 = \delta(t) \cdot \sum_{m,n}|nm\rangle\langle nm| = {\bf 1}.
\ee 
The trace of a local operator $\mathcal O$ over the Hilbert space and time is defined as follows
\be
\Tr(\mathcal O) \equiv \sum_{n,l,t}\langle nlt|\mathcal O|nlt\rangle=\int dt \sum_{n,l} \langle nl|\mathcal O(t)|nl\rangle\,.
\ee

\subsection{Setting up the ``Feynman rules''}
In this Section we will derive the differential operators that will appear in the vertices of the diagrams in Fig. \ref{Diagrams}. 

First, we expand the classical action to the second order in {\it external} electromagnetic field 
\be
S_{nr}= S_{nr}^{(0)} + S_{nr}^{(1)} + S_{nr}^{(2)}\,.
\ee
The unperturbed action is given by
\bea\nonumber
S_{nr}^{(0)}&&=\int d^2xdt \,\,\psi^\dag\left[i\p_0-\omega_c\left(a^\dag a + \frac{1}{2}\right)\right]\psi \\
&&= \int d^2xdt \,\,\Psi^\dag G_0^{-1}\Psi \,.
\eea
The part of the action linear in external fields is given by
\bea\nonumber
\!\!\!S_{nr}^{(1)} &=& \int d^2xdt \, \psi^\dag\Big[\delta A_0 - \frac{1}{2\sqrt{2}m\ell}(\{a^\dag,\delta A_{\bar{z}}\}+\{a,\delta A_{z}\})\Big]\psi \\
&=& \int d^2x dt \,\,\psi^\dag V^{(1)} \psi\,,
\eea
where $\{a, \delta A_z\}$ is the anticommutator (recall that $a$ is a differential operator that we understand as acting to the right).

The part of the action quadratic in external fields is given by
\be
 S_{nr}^{(2)} = -\int d^2xdt\,\, \psi^\dag\left[\frac{1}{2m}|\delta A|^2\right]\psi =  \int d^2x dt\,\, \psi^\dag V^{(2)} \psi\, .
\ee
The full ``vertex operator'' consists of the terms linear and quadratic in external fields
\be
V(x,t) = V^{(1)}(x,t) + V^{(2)}(x,t)\,.
\ee
Using \eqref{zab} we interpret $V$ as an operator on the Fock space.

We re-write all the vertices in Fourier space and introduce a vector $\mathcal V^{(1)}_\mu({\mathbf{k},\Omega})$ according to
\be
V^{(1)} = \mathcal V^{(1)}_\mu({\mathbf{k}},\Omega) \delta A_\mu({\mathbf{k}},\Omega)\, ,
\ee
which is always possible because $V^{(1)}$ is linear in the external fields by definition. Consider the terms in $V$ linear in, say,  $\delta A_{\bar z}$
\be
V(x,t)\Big|_{\delta A_{\bar z}}  =  \frac{1}{2\sqrt{2}m\ell}\{a^\dag,\delta A_{\bar{z}}\}\,.
\ee
In momentum space this takes form
\be
 V({\mathbf{k}},\Omega)\Big|_{\delta A_{\bar z}}  = e^{-i\Omega t}\frac{1}{2\sqrt{2}m\ell}\{a^\dag,e^{i{\mathbf{k}\cdot \mathbf{x}}}\}\delta A_{\bar{z}}({\mathbf{k}},\Omega)\,.
\ee
Then, using \eqref{zab}
\be\la{exponent}
e^{i\vec{k}\vec{x}} = e^{\frac{i}{2}k\bar{z}}e^{\frac{i}{2}\bar{k}z} = e^{-\frac{k\ell}{\sqrt{2}}a^\dag}e^{\frac{\bar{k}\ell}{\sqrt{2}}a}e^{-\frac{k\ell}{\sqrt{2}}b}e^{\frac{\bar{k}\ell}{\sqrt{2}}b^\dag}\,,
\ee
where we introduced the complex momentum $k = k_1 + ik_2$.
Finally, using that $a$'s and $b$'s commute with each other we get an expression for
\be\la{OMG-2}
 \mathcal V^{(1)}_{ z}({\mathbf{k},\Omega}) = \frac{1}{2\sqrt{2}m\ell}e^{-i\omega t}e^{-\frac{k\ell}{\sqrt{2}}b}e^{\frac{\bar{k}\ell}{\sqrt{2}}b^\dag}\Big\{a^\dag, e^{-\frac{k\ell}{\sqrt{2}}a^\dag}e^{\frac{\bar{k}\ell}{\sqrt{2}}a}\Big\}\,.
\ee
Expressions for the other vertices can be derived in the same way
\bea\la{OMG-0}
&&\!\!\!\!\!\!\!\mathcal{V}^{(1)}_0 = e^{-i{\Omega} t}e^{-\frac{k\ell}{\sqrt{2}}b}e^{\frac{\bar{k}\ell}{\sqrt{2}}b^\dag}e^{-\frac{k\ell}{\sqrt{2}}a^\dag}e^{\frac{\bar{k}\ell}{\sqrt{2}}a}\,,\\
&&\!\!\!\!\!\!\!\mathcal{V}^{(1)}_{\bar z}= \frac{-1}{2\sqrt{2}m\ell}e^{-i{\Omega} t}e^{-\frac{k\ell}{\sqrt{2}}b}e^{\frac{\bar{k}\ell}{\sqrt{2}}b^\dag}\{a^\dag,e^{-\frac{k\ell}{\sqrt{2}}a^\dag}e^{\frac{\bar{k}\ell}{\sqrt{2}}a}\}\,.
\eea
Notice that part of the vertices that depends on both time and $b$'s has completely factorized and {\it is the same for all vertices}. We will be able to use this fact to integrate over time and to trace over the Fock space generated by $b^\dag$ before tracing over the Fock space generated by $a^\dag$. It is the trace over $a$ where all of the complexity is concentrated. For this reason it will be convenient to introduce a separate notation for the part of the ``vertex operators'' that acts only in the Fock space generated by $a^\dag$. Thus we define
\be
\mathcal V^{(1)}_\mu = e^{-i{\Omega} t}e^{-\frac{k\ell}{\sqrt{2}}b}e^{\frac{\bar{k}\ell}{\sqrt{2}}b^\dag} \tilde{V}_\mu\,,
\ee
where
\bea
\la{V0}
\tilde{V}_0 &=& 1\,, \\ 
\tilde{V}_{\bar z} &=& -\frac{1}{2\sqrt{2}m\ell}\{a^\dag,e^{-\frac{k\ell}{\sqrt{2}}a^\dag}e^{\frac{\bar{k}\ell}{\sqrt{2}}a}\}\,, \\\la{V2}
\tilde{V}_z &=& -\frac{1}{2\sqrt{2}m\ell}\{a,e^{-\frac{k\ell}{\sqrt{2}}a^\dag}e^{\frac{\bar{k}\ell}{\sqrt{2}}a}\}\,.
\eea

\subsection{Generating Functional to the Second Order}
In this Section we will perform an {\it exact} computation of the entire quadratic generating functional to {\it all} orders in the gradient expansion. Before diving into details we briefly pause to mention a few relations that will be heavily used in the sequel
\bea
&&[b,f(b^\dag)] = f^\prime(b^\dag)\,,\\
&&e^{Qb}f(b^\dag) = f(b^\dag + Q) e^{Qb}\,.
\eea
Using these relations and elementary properties of the oscillator algebra we can evaluate the following expectation values
\bea\la{Lag-a}
\!\!\!\!\!\!\!\langle n| e^{-\frac{k\ell}{\sqrt{2}}a^\dag}e^{\frac{\bar{k}\ell}{\sqrt{2}}a}|m \rangle &=& \!\!\sqrt{\frac{n!}{m!}}\left(\frac{\bar{k}\ell}{\sqrt{2}}\right)^{m-n}L^{m-n}_n\left(\!\frac{|k\ell|^2}{2}\!\right),\\
&=&\!\!\sqrt{\frac{m!}{n!}}\left(\frac{-k\ell}{\sqrt{2}}\right)^{n-m}L^{n-m}_m\left(\!\frac{|k\ell|^2}{2}\!\right).
\eea
Similar equations can be found in [\onlinecite{perelomov1977generalized}].

There will be two major contributions to the generating functional in quadratic order. One contribution comes from the contact terms. These are obtained by plugging $V^{(2)}$ into
\be
-i \Tr G_0 V\,.
\ee
These contributions are always evaluated at zero momentum and zero frequency. In fact, the contact terms can be restored simply via analyzing the Ward identities for electric charge conservation. We will denote the contribution of the contact terms to the polarization operator via $\Pi^{\mu\nu}_{c, \rm nr}$.

 The main contribution comes from
\be
\frac{i}{2}\Tr G_0V^{(1)} G_0 V^{(1)} \,.
\ee
First, we will trace over the  Fock space generated by $b, b^\dag$, then over frequency and in the end we will be left with an irreducible expression for the trace over the Fock space generated by $a, a^\dag$. The details of the steps outlined above can be found in the Appendix \ref{Derivation}. 
 
We find
\begin{widetext} 
\bea\nonumber
 &&\!\!\!\!\!\Tr G_0V^{(1)} G_0 V^{(1)}  = \sum_{n,l,t}\langle tnl | G_0V^{(1)}G_0V^{(1)} |tnl\rangle \\ \la{diagramanswer}
 &&\!\!\!\!\!\!\!\!\!\!= \frac{m}{4\pi}\int \frac{d^2k d\Omega }{(2\pi)^3}e^{-\frac{|k\ell|^2}{2}}\!\!\!\!\!\!\sum_{n^\prime < N, n \geq N}\!\!\!\!\!\! \frac{\langle n | \tilde V_\mu^{(1)}(k)| n^\prime \rangle\langle n^\prime | \tilde V_\nu^{(1)}(-k)| n\rangle +\langle n^\prime | \tilde V_\nu^{(1)}(k)| n\rangle\langle n | \tilde V_\mu^{(1)}(-k)| n^\prime\rangle }{E_{n} - E_{n^\prime} - \Omega} \delta A_\mu(k)\delta A_\nu(-k).
\eea
In the remainder of the Section we will simplify this expression.

We introduce the following notation
\be
\Gamma^\mu_{n n^\prime}(k,\Omega) = \langle n | \tilde V^{(1)}_\mu(k)| n^\prime \rangle\,,
\ee
then (using the dimensionless frequency $\omega = \Omega/ \omega_c$)
\be\la{W-answer}
\Pi_{\rm nr}^{\mu\nu}(k,\omega) = \frac{i}{(4\pi \ell^2)}\frac{1}{\omega_c} e^{-\frac{|k\ell|^2}{2}}\sum_{n^\prime< N,n\geq N}\frac{\Gamma^\mu_{nn^\prime}(k)\Gamma^\nu_{n^\prime n}(-k)+\Gamma^\nu_{nn^\prime}(k)\Gamma^\mu_{n^\prime n}(-k)}{n-n^\prime-\omega} + \Pi^{\mu\nu}_{c, \rm nr}\,.
\ee
This is the main result of the Section. In the following we will show that all of the components of the polarization tensor can be reconstructed from a {\it single} generating function.

\end{widetext} 

\subsection{The Generating Function}
While (\ref{W-answer}) is indeed the final expression that cannot be reduced further, it is not convenient to work with since one has to use complicated expressions for $\Gamma^\mu_{nn^\prime}$. We will introduce a trick that will allow to express {\it all} of the components of the polarization operator in terms of derivatives of a {\it single} function. 

We define the generating function $\mathcal G(k,k^\prime;N)$
  \begin{widetext} 
\bea
 \mathcal G(k,k^\prime;N)&&=\sum_{n\geq N, n^\prime< N} \left(\frac{\Gamma^0_{nn^\prime}(k)\Gamma^0_{n^\prime n}(k^\prime)}{n-n^\prime-\omega}+ \frac{\Gamma^0_{nn^\prime}(k^\prime)\Gamma^0_{n^\prime n}(k)}{n-n^\prime+\omega} \right) 
 \\ &&= \sum_{n\geq N,n^\prime< N}\left(-\frac{\ell^2}{2}\right)^{n-n^\prime}\frac{n^\prime!}{n!}\left(\frac{(k\bar{k}^\prime)^{n-n^\prime}}{n-n^\prime-\omega}+\frac{(\bar{k}k^\prime)^{n-n^\prime}}{n-n^\prime+\omega}\right)  L_{n^\prime}^{n-n^\prime}\left(\frac{|k\ell|^2}{2}\right)L_{n^\prime}^{n-n^\prime}\left(\frac{|k^\prime\ell|^2}{2}\right)\,.
\eea
\end{widetext} 
First, we notice (with the help of \eqref{Lag-a}) that
\be
\Pi_{nr}^{00} =  \frac{m}{4\pi}e^{-\frac{|k\ell|^2}{2}}\mathcal G(k,-k;N)\,.
\ee
other components of $\Pi^{\mu\nu}_{\rm nr}$ can be expressed as derivatives of $\mathcal G(k,k^\prime;N)$ with respect to momenta. To see this we use the identities
\bea
e^{-ka^\dag}e^{\bar{k}a}a^\dag &=& (-\p_k+\bar{k})e^{-ka^\dag}e^{\bar{k}a}\,,\\
ae^{-ka^\dag}e^{\bar{k}a} &=&(\p_{\bar{k}}-k)e^{-ka^\dag}e^{\bar{k}a}\, .
\eea
These identities allow us to re-write the vertex insertions (\ref{V0})-(\ref{V2}) in terms of derivatives with respect to momentum as follows 
\bea\la{V0k}
\!\!\!\!\!\tilde V_0 (k) &=& e^{-\frac{k\ell}{\sqrt{2}}a^\dag}e^{\frac{\bar{k}\ell}{\sqrt{2}}a}\,,\\
\!\!\!\!\!\tilde V_{\bar z}(k) &=& -\frac{1}{2\sqrt{2}\ell m}\left(-\frac{2\sqrt{2}}{\ell}\p_k+\frac{\ell}{\sqrt{2}}\bar{k}\right) e^{-\frac{k\ell}{\sqrt{2}}a^\dag}e^{\frac{\bar{k}\ell}{\sqrt{2}}a}\,,  \\ \la{Vzk}
\!\!\!\!\!\tilde V_z(k) &=&-\frac{1}{2\sqrt{2}\ell m}\left(\frac{2\sqrt{2}}{\ell}\p_{\bar{k}} -\frac{\ell}{\sqrt{2}}k\right) e^{-\frac{k\ell}{\sqrt{2}}a^\dag}e^{\frac{\bar{k}\ell}{\sqrt{2}}a}\,.
\eea
Next we introduce a separate notation for the differential operators acting on $e^{-\frac{k\ell}{\sqrt{2}}a^\dag}e^{\frac{\bar{k}\ell}{\sqrt{2}}a}$ in Eqs.\eqref{V0k}-\eqref{Vzk} as follows
\bea\la{P0k}
\!\!\!\!\!\hat{\mathcal P}^0_{\rm nr}(k) &=& 1\,,\\
\!\!\!\!\! \hat{\mathcal P}^{ z}_{\rm nr}(k) &=& -\frac{1}{2\sqrt{2}\ell m}\left(-\frac{2\sqrt{2}}{\ell}\p_k+\frac{\ell}{\sqrt{2}}\bar{k}\right)\,,  \\ \la{Pzk}
\!\!\!\!\! \hat{\mathcal P}^{\bar z}_{\rm nr}(k) &=&-\frac{1}{2\sqrt{2}\ell m}\left(\frac{2\sqrt{2}}{\ell}\p_{\bar{k}} -\frac{\ell}{\sqrt{2}}k\right) \,.
\eea

Then an arbitrary element of the polarization operator is given by
\be\la{Polanswer}
\Pi_{nr}^{\mu\nu}(\omega,k) =\frac{m}{4\pi}e^{-\frac{|k\ell|^2}{2}}\lim_{k^\prime\rightarrow -k}\hat{\mathcal P}^\mu_{\rm nr}(k)\hat{\mathcal P}^\nu_{\rm nr}(k^\prime)\mathcal G(k,k^\prime;N)+\Pi^{\mu\nu}_{c, \rm nr}\,.
\ee
This expression is the one we will use for practical computations and is the first main result of the present manuscript.

The contact terms are obtained from $\Tr G_0 V^{(2)}$. The only contact term is the well-known diamagnetic term given by
\bea\nonumber
W^{(2)}_{c, \rm nr} = -i \Tr G_0V^{(2)} =N\frac{\omega_c}{4\pi} \int \frac{d^2k}{(2\pi)^2}\delta A_z(k)\delta A_{\bar z}(-k)\, .
\eea
This term is evaluated at zero frequency $\omega=0$. The contribution to momentum space polarization tensor is
\be\la{Polcontact}
\Pi^{z\bar z}_{c, \rm nr} =N\frac{\omega_c}{16\pi} \delta_{z\bar z}\,.
\ee
It can be checked explicitly that this term restores the Ward identity \eqref{gaugeWI}.
Direct application of Eqs. \eqref{Polanswer}-\eqref{Polcontact} to $N$ filled Landau levels allows us to write the generating functional in the leading orders in the gradient expansion
\bea
&&W[A_\mu] =\frac{N}{4\pi} \int d^2x dt \Big[ Ad  A  \\
&&+\frac{1}{ \omega_c} | \delta \vec E|^2-\frac{N}{m} \delta B^2 - \frac{3N}{2} \ell^2  \delta B (\p_i \delta E_{i}) +\ldots\Big]\,,
\eea
where $E_i$ and $B$ are made from the perturbations of the electromagnetic field $\delta A_\mu$. We have also absorbed the linear term $\bar \rho \delta A_0$ into the Chern-Simons term by including the background $\bar A_\mu$. Higher order terms can also be easily obtained from \eqref{Polanswer}. Finally, note that $| \delta \vec E|^2$ and $B^2$ terms do {\it not} combine into $\delta F^{\mu \nu} \delta F_{\mu \nu}$ due to apparent absence of Lorentz invariance. We have also checked that Ward Identities of the Galilean symmetry studied in \onlinecite{hoyos2012hall} are satisfied.
\subsection{Including the $\mathrm g$-factor}
For the future applications we also need to include the effects of finite $\mathrm g$-factor of the electron, $\rg$, by adding an extra ``Zeeman'' term to the matter action
\be
\delta S_{nr}[\psi, \psi^\dag] =  \frac{\rg}{4m}\int B\psi^\dag \psi\,.
\ee
This results in the redefinition of the number current 
\be\la{current-g}
J^{i}(g) = J^{i}(0) + \frac{\rg}{4m} \epsilon^{ij} \p_j \rho
\ee
as well as the vertices
\bea\la{P0kg}
\!\!\!\!\! \hat{\mathcal P}^{ z}_{\rm nr}(k;g) &=& -\frac{1}{2\sqrt{2}\ell m}\left(-\frac{2\sqrt{2}}{\ell}\p_k+\left(1-\frac{\rg}{2}\right)\frac{\ell}{\sqrt{2}}\bar{k}\right)\,,  \\ \la{Pzk}
\!\!\!\!\! \hat{\mathcal P}^{\bar z}_{\rm nr}(k;g) &=&-\frac{1}{2\sqrt{2}\ell m}\left(\frac{2\sqrt{2}}{\ell}\p_{\bar{k}} -\left(1-\frac{\rg}{2}\right)\frac{\ell}{\sqrt{2}}k\right) \,.
\eea
Note that the generating functional for finite $\rg$ can be expressed in terms of the generating functional at $\rg=0$ as follows
\be\la{finiteg}
W_{\rm nr}\left[\delta A_0,\delta A_i; \rg\right] =W_{\rm nr}\left[\delta A_0+ \frac{\rg}{4m} \delta B,\delta A_i; \rg=0\right]\,.
\ee
In particular, Eq.\eqref{finiteg} implies the following relations between the components of the polarization tensor 
\bea\la{Pi00g}
\Pi^{00}_{\rm nr}(\mathbf{k},\Omega;\rg) &=& \Pi^{00}_{\rm nr}(\mathbf{k},\Omega;0) \\
\la{Pi0ig}
\Pi^{0i}_{\rm nr}(\mathbf{k},\Omega;\rg)&=&\Pi^{0i}_{\rm nr}+  i \frac {\rg}{4m} \epsilon^{ij}k_j \Pi_{\rm nr}^{00}(\mathbf{k},\Omega;0) \\
\nonumber
\Pi^{ij}_{\rm nr}(\mathbf{k},\Omega;\rg)&=&\Pi^{ij}_{\rm nr}(\mathbf k, \Omega; 0) \\ \nonumber
&&\!\!\!\!\!\!\!\!\!\!\!\!+ i\frac{\rg}{4m}k_l \Big(\epsilon^{lj} \Pi^{0i}_{\rm nr}(\mathbf k, \Omega; 0)-\epsilon^{li}\Pi^{j0}_{\rm nr}(\mathbf k, \Omega; 0)\Big) \\ \la{Piijg}
&&\!\!\!\!\!\!\!\!\!\!\!\!+ \frac{\rg^2}{16 m^2} \left(|k|^2 \delta^{ij} - k^i k^j\right) \Pi^{00}_{\rm nr}\left(\mathbf k, \Omega; 0\right) \,.
\eea

The $\mathrm g$-factor will be used in Section IV to take the LLL projection and as an extra control parameter in Section V where we will compare the large $N$ limit of the  non-relativistic polarization operator with a semiclassical computation.
\section{Dirac electrons}
\label{sec:PolDirac}
In this Section we calculate the electromagnetic response of the Dirac electrons filling $N$ Landau levels.
\subsection{Model and the ``Feynman rules''}

In this Section we will calculate the polarization tensor for Dirac\footnote{\protect{The Dirac nature of the fermions does not have to come from spacetime symmetries. It can be also internal $SU(2)$ symmetry as it happens in graphene.}} fermions in strong magnetic field, filling $N$ Landau levels. The action is given by
\be\la{Diracaction}
S_D = \int d^3x \bar \Psi \slashed D \Psi\,,
\ee
where $\Psi$ is a two-component spinor,  $\bar \Psi = \Psi^\dag \gamma^0$ and $\slashed D = \gamma^0 D_0+v_F \gamma^i D_i$, where the covariant derivative $D_\mu$ is given by \eqref{CovDerivative} and includes both constant magnetic field and its perturbations, $v_F$ is the Fermi velocity. To emphasize Lorentz invariance we will use the notation $d^3x = dt d^2x$.

The Hamiltonian is
\begin{equation}
H=-iv_F\gamma^0\gamma^iD_i-A_0\,,
\end{equation}  
where $v_F$ is the Fermi velocity.
We choose the $\gamma$-matrices as follows
\begin{equation}
\gamma^0=\sigma^3, \qquad \gamma^1=i\sigma^2, \qquad \gamma^2=-i\sigma^1,
\end{equation}
where $\sigma^i$ are the Pauli's matrices.

As in the non-relativistic case the Hilbert space maps on two copies of the Fock space generated by $a^\dag$ and $b^\dag$ defined in \eqref{adef}-\eqref{zab}. The unperturbed Hamiltonian can be explicitly written as 
\begin{equation}
H_0 = \begin{pmatrix}
-\mu & -v_F\sqrt{2\bar B} a^\dagger \\
-v_F\sqrt{2\bar B}a & -\mu
\end{pmatrix},
\end{equation}
where $\mu$ is the chemical potential. There are three types of eigenstates of $H$

\par (i) zero energy, $\ket{\Psi_{0,m}} = \begin{pmatrix} \ket{0,m} \\ 0 \end{pmatrix},~ E^D_{0} +\mu = 0$,

\par  (ii) positive energy, $\ket{\Psi^+_{n,m}} =\frac{1}{\sqrt{2}} \begin{pmatrix} \ket{n,m} \\ -\ket{n-1,m}\end{pmatrix},\\ E^{D+}_{n} + \mu = +v_F\sqrt{2\bar B n}$.

\par (iii) negative energy, $\ket{\Psi^-_{n,m}} = \frac{1}{\sqrt{2}}\begin{pmatrix} \ket{n,m} \\ +\ket{n-1,m}\end{pmatrix},\\ E^{D-}_{n} + \mu = -v_F\sqrt{2\bar B n}$.

We introduce a uniform notation for all of the eigenstates as follows
\begin{align}
\ket{\Psi_{n,m}} &= \text{norm}(n)\begin{pmatrix}
\ket{|n|,m} \\ -\sign(n) \ket{(|n|-1),m}
\end{pmatrix}=\ket{\Psi_n}  \otimes \ket{m}\,,\nonumber\\
E^D_{n} &= \sign(n) v_F \sqrt{2\bar B|n|} - \mu,\la{psinm}
\end{align}
where it is understood that $\ket{-1,m} = 0$ and $ \text{norm}(n)=1/\sqrt{2}$ for $|n|>0$ and $ \text{norm}(0)=1$. With this notation at hand $n \in \mathbb{Z}$, $m \in \mathbb Z_+$. The unperturbed Green's function is
\be
	G_0(t)  = \int \frac{d\Omega}{2\pi} e^{-i \Omega t}\sum_{n ,m} \frac{\ket{\Psi_{n,m}}\bra{\Psi_{n,m}}}{\Omega - E^D_{n} + i \e ~\sign(E^D_{n})}\,.\\
\ee

The massless Dirac action is easily decomposed into the terms free of and linear in external electromagnetic field perturbations (about the constant magnetic field and chemical potential),
\be
S_D = S_D^{(0)} + S_D^{(1)}\,,
\ee
where
\begin{align}
S_D^{(0)} =& \int d^3x  \Psi^\dagger
\begin{pmatrix}
i \d_0 + \mu & v_F\sqrt{2\bar B} a^\dagger \\
v_F\sqrt{2\bar B} a & i \d_0 + \mu
\end{pmatrix}
\Psi,\nonumber\\
S_D^{(1)} =& \int d^3x \Psi^\dagger 
\begin{pmatrix}
\delta A_0 & v_F \delta A_z \\
v_F\delta A_\zb & \delta A_0
\end{pmatrix}
\Psi.
\end{align}
The first term gives the bare propagator,
\begin{align}
S^{(0)}_D = \int d^3x \Psi^\dagger G^{-1}_0 \Psi,
\end{align}
that satisfies
\be
G_0^{-1} |\Psi_{n,m} \rangle= (\Omega+\mu -E^D_n) |\Psi_{n,m}\rangle\,,
\ee
and
\be
 G_0^{-1}G_0 = \delta(t) \sum_i {\ket{\Psi_i}\bra{\Psi_i}} = \mathds{1}\,.
\ee
The second term gives the vertex in position space 
\begin{align}
V(t,x) = \begin{pmatrix}
\delta A_0(t,x) & v_F\delta A_z(t,x) \\
v_F \delta A_\zb(t,x) & \delta A_0(t,x)
\end{pmatrix}.
\end{align}
Note that the vertices have no explicit coordinate dependence or derivatives, and it simply remains to Fourier transform them, using  \eqref{zab} and \eqref{exponent}.

Following the Section \ref{sec:method}, we wish to evaluate $\Tr G_0 V G_0 V$ to derive the generating functional. In this case there are no contact terms because Dirac Hamiltonian (and the action) is linear in external fields. Using the results of Appendices \ref{Derivation} and \ref{cohstates} we can take the trace over the Fock space generated by $b^\dag$ and over time 
	\begin{align}
	&\Tr G_0 V G_0 V=\int dt \sum_{m,n} \bra{\Psi_{n,m}(t)}G_0 V G_0 V \ket{\Psi_{n,m}(t)}\nonumber\\
	&=\frac{1}{4 \pi \ell^2}\int \frac{d\Omega'}{2\pi}\frac{d\Omega}{2\pi}\frac{d^2\mathbf{k}}{(2\pi)^2}e^{-\frac{|k\ell|^2}{2}}\nonumber\\
	&\times\sum_{nn'}\frac{\hat{V}_{nn'}(\mathbf{k},\Omega)}{\Omega'+\Omega-E^D_n+i\epsilon  \sign(E^D_n)}\frac{\hat{V}_{n'n}(-\mathbf{k},-\Omega)}{\Omega'-E^D_{n'}+i\epsilon \sign{(E^D_{n'})}},
	\end{align}
in which the vertex operator in the momentum space is defined as
	\begin{align}
	\hat{V}_{nn'}(\mathbf{k},\Omega) =\bra{\Psi_n} \begin{pmatrix}
	\delta A_0(-\mathbf{k},-\Omega) & v_F\delta A_z(-\mathbf{k},-\Omega) \\
	v_F \delta A_\zb(-\mathbf{k},-\Omega) & \delta A_0(-\mathbf{k},-\Omega)
	\end{pmatrix}\nonumber\\\times e^{-\frac{k\ell a^\dagger}{\sqrt{2}}}e^{-\frac{\bar{k}\ell a}{\sqrt{2}}}\ket{\Psi_{n'}}.
	\end{align}

The $i\epsilon$ prescription  is crucial in evaluating the frequency integral
\begin{align}
&\int \frac{d\Omega'}{2\pi} \frac{1}{\Omega'+\W - E^{D}_n + i \e ~ \sign(E^{D}_n)}  \frac{1}{\Omega' - E^{D}_{n'} + i \e ~ \sign(E^{D}_{n'})} \nonumber \\
&= 
\left\{\begin{array}{ccc}
\frac{i}{E^{D}_n - E^D_{n'} - \W} & , & E^{D}_n < 0,~E^{D}_{n'}>0 \\
\frac{i}{E^{D}_{n'} - E^{D}_n + \W}  & , & E^{D}_n>0,~E^{D}_{n'}<0 \\
0 & , & \mathrm{else}\end{array}\right. \,.
\end{align}
The polarization tensor $\Pi_D^{\mu\nu}(\Omega,\mathbf{k})$ is given by
\bea
\nonumber
&&\Pi_D^{\mu\nu}(\Omega,\mathbf{k})=-\frac{e^{-\frac{|k\ell|^2}{2}}}{2\pi \ell^2}\sum_{n'\le N,n>N }\left(\frac{\Gamma^\mu_{Dnn'}(\mathbf{k})\Gamma^\nu_{Dn'n}(-\mathbf{k})}{E^D_{n}-E^D_{n'}-\Omega}\right.\\\label{eq:polDirac}
&&\left.+\frac{\Gamma^\nu_{Dnn'}(-\mathbf{k})\Gamma^\mu_{Dn'n}(\mathbf{k})}{E^D_{n}-E^D_{n'}+\Omega} \right),
\eea
where 
\begin{equation}\la{GammaD}
\Gamma^\mu_{Dnn'}(\vec{k})=\bra{\Psi_n} \mathcal P_D^\mu e^{-\frac{k\ell a^\dagger}{\sqrt{2}}}e^{-\frac{\bar{k}\ell a}{\sqrt{2}}}\ket{\Psi_{n'}}
\end{equation} 
and
\begin{eqnarray}
\mathcal P_D^0=\mathds{1}, \quad \mathcal P_D^1=v_F\sigma^1, \quad  \mathcal P_D^2=v_F\sigma^2. 
\end{eqnarray}

Using \eqref{Lag-a} we evaluate the vertices $\Gamma^\mu_{Dnn'}(\mathbf{k})$. The expressions turn out to be quite complicated and so we list them in the Appendix D.

\subsection{Dirac Polarization Tensor}
 In this Section, write out explicit expressions for the Landau level polarization tensor for Dirac fermion in the leading order in momentum and frequency. While Eq.(\ref{eq:polDirac}) looks similar to the corresponding Eq.\eqref{diagramanswer} for the non-relativistic fermions we want to emphasize that there is a difficulty in evaluating the summation, even when we limit ourselves to some finite order in momentum and frequency. The reason is that every component of the polarization tensor, contains the sum over $n'$ ({\it i.e.} the sum over the Dirac sea) from $-\infty$ to $N$, where $N$ is the number of filled Landau levels. We remind the reader that in the non-relativistic case this sum consisted of a {\it finite} number of terms (because the parabolic dispersion relation has a bottom, see Fig. \ref{diractransitions}) in every order in momentum and frequency. In the present case the the sum has infinite number of terms, however, it is {\it convergent} and does {\it not} need to be regularized. 
 To simplify the expressions we fix a coordinate frame in which $\mathbf{k}=(k_1,0)$. Leaving the details to the Appendix \ref{sec:DiracEx} we present the leading order terms below
\begin{widetext}
	\begin{align}
	\Pi^{12}_D(\Omega,\mathbf{k})&=i\Omega \frac{N+1/2}{2\pi}-i\Omega (k_1\ell)^2 \frac{6N^2+6N+1}{16\pi}+i\Omega^3 \frac{\ell^2}{v_F^2}\frac{8N^2+8N+1}{8\pi}+\cdots\\
	\Pi^{00}_D(\Omega,\mathbf{k})&=-k_1^2\frac{3\ell}{2\sqrt{2}\pi v_F}\zeta\left(-\frac{1}{2},N+1\right)+\cdots\\
	\Pi^{11}_D(\Omega,\mathbf{k})&=-\Omega^2\frac{3\ell}{2\sqrt{2}\pi v_F}\zeta\left(-\frac{1}{2},N+1\right)+\cdots\\
	\Pi^{22}_D(\Omega,\mathbf{k})&=-\Omega^2\frac{3\ell}{2\sqrt{2}\pi v_F}\zeta\left(-\frac{1}{2},N+1\right)+k_1^2\frac{3\ell v_F}{4\sqrt{2}\pi }\zeta\left(-\frac{1}{2},N+1\right)+\cdots\,,
	\end{align}	
where $\zeta(s,n)$ is the Hurwitz $\zeta$-function \footnote{See the appendix \ref{sec:DiracEx} for more detail}. We stress that each component of the polarization tensor is {\it finite} without any need for regularization. The use of $\zeta$-function is a convenient choice that allows to evaluate the sums analytically.

In the coordinate space the generating functional is given by
\be\la{GenDirac}
W_{\rm D} = \int d^3 x \left[ \frac{N+ \frac{1}{2}}{4\pi} AdA - \frac{3\ell}{4\sqrt{2}\pi v_F}\zeta\left(-\frac{1}{2},N+1\right) |\delta\vec{E}|^2 + \frac{3\ell v_F}{8\sqrt{2}\pi }\zeta\left(-\frac{1}{2},N+1\right) \delta B^2 - \ell^2 \frac{6(N+ \frac{1}{2})^2-\frac{1}{2}}{8\pi} \delta B (\p_i \delta E_i)  + \ldots\right]\,.
\ee
Eq.\eqref{GenDirac} is valid in arbitrary coordinate frame. Eq. \eqref{GenDirac} the main result of the present Section. Note that despite the Lorentz invariance of the action \eqref{Diracaction} the generating functional is not Lorentz invariant. This happens because the the Lorentz invariance is broken by the background magnetic field, which is held at a fixed value.
\end{widetext}	

We can subject the above results to several checks. First, define the finite frequency and momentum corrections to the Hall conductivity via
\be
\sigma^{12}(\Omega,\mathbf{k})= \sigma^H(0) + \sigma^H_{k^2}|\mathbf{k}\ell|^2  +  \sigma^H_{\Omega^2}\Omega^2 +\ldots
\ee
According to the Ref. \onlinecite{GRS} there is a relations between $\Omega^2$ and $|\mathbf k\ell|^2$ coefficients of finite frequency and momentum Hall conductivity defined via 
\be\la{Dirack2}
2\sigma^{H}_{k^2}+v_F^2\sigma^{H}_{\Omega^2}=\frac{\mathcal S\ell^2}{4\pi}\,,
\ee
where $\mathcal S=N(N+1)$ is the relativistic version of the Shift \cite{wen1992shift} of the integer quantum Hall state of Dirac electrons at filling fraction $\nu=N$. We can check explicitly that Eq. \eqref{Dirack2} holds.

 Next, we compare the polarization tensors for the Dirac and non-relativistic electrons in the large $N$ limit. Using the results of Section \ref{sec:method} we have in the non-relativistic case
\begin{align}
\Pi^{12}_{\text{nr}}(\Omega,\mathbf{k})&=i\Omega \frac{N}{2\pi}-i\Omega k_1^2 \ell^2 \frac{3N^2}{8\pi}+i\Omega^3\ell^2\frac{N^2}{\pi v_F^2}+\cdots\,,\\
\Pi^{00}_{\text{nr}}(\Omega,\mathbf{k})&=k_1^2\frac{\ell N^{3/2}}{v_F\sqrt{2}\pi}+\cdots\,,\\
\Pi^{11}_{\text{nr}}(\Omega,\mathbf{k})&=\Omega^2\frac{\ell N^{3/2}}{v_F\sqrt{2}\pi}+\cdots\,,\\
\Pi^{22}_{\text{nr}}(\Omega,\mathbf{k})&=\Omega^2\frac{\ell N^{3/2}}{v_F\sqrt{2}\pi}-k_1^2\frac{ \ell v_F N^{3/2}}{2\sqrt{2}\pi}\cdots\,,
\end{align}	
 where $\omega_c$ can be written in terms of Fermi velocity $v_F$ and Fermi momentum $k_F$ as follows
\begin{equation}
\omega_c=\frac{\bar B}{m}=\frac{\bar B v_F}{k_F}=\frac{v_F\sqrt{\bar B}}{\sqrt{2N}}\,,
\end{equation}
where we used the relation between filling fraction and Fermi momentum 
\begin{equation}
N=\frac{\bar \rho}{\bar B/2\pi}=\frac{k_F^2}{2\bar B},
\end{equation}
where $\bar \rho$ is the non-relativistic electron density. 

Using the asymptotic formula for the $\zeta$-function at large $N$ 
\begin{equation}
\zeta\left(-\frac{1}{2},N+1\right)\approx-\frac{2}{3}\left(N+\frac{1}{2}\right)^{3/2}
\end{equation}
we find that the non-relativistic and Dirac polarization tensors agree in leading and sub-leading order in $N$, provided we replace $N\longrightarrow N+\frac{1}{2}$. The latter replacement comes up due to the contribution of the Berry phase in the Dirac's case. We stress that the equivalence holds when the $\mathrm g$-factor of the non-relativistic electrons vanishes and does {\it not} equal to $2$ as one may naively expect.

The terms that are sub-sub-leading order in $N$ do not agree, which can be shown by an explicit calculation. The agreement of the leading and sub-leading orders is not surprising, since in large $N$ limit, which is the case of high density and small applied magnetic field, the semiclassical approximation applies equally well to both systems, however Dirac theory has an extra Berry phase contribution. We will study the large $N$ limit in more detail in the Section V.
\section{Universality of the Lowest Landau Level}
\la{DvsNR}

In this Section we will show that the exact electromagnetic linear response functions of the lowest Landau level of non-relativistic and Dirac electrons agree to all orders in the gradient expansion. Note that LLL means $N=1$ for the non-relativistic case and $N=0$ for Dirac.

It was demonstrated in Ref.[\onlinecite{Son:2015xqa}] that the non-relativistic action \eqref{NR-action} reduces to  \eqref{Diracaction} in the limit $m\rightarrow 0$, $~\rg=2$ and provided that transitions across the gap are neglected. This argument was used to deduce the following relationship between the generating functionals
\be
W_{\rm D}[\delta A_\mu] = W_{\rm nr}[\delta A_\mu] - \frac{1}{2} \frac{1}{4\pi} \int AdA\,.
\ee
This relation holds only in the leading order in the gradient expansion. We will show that there is an exact version of this relation which reads
\be\la{DvsNRW}
W_{\rm D}[\delta A_\mu] = W_{\rm nr}[\delta A_\mu] - \frac{1}{2}W_{\rm nr}[\delta A_\mu] = \frac{1}{2}W_{\rm nr}[\delta A_\mu] \,.
\ee
Eq. \eqref{DvsNRW} can be understood as follows. Completely filled $0$-th Landau level contributes $W_{nr}[\delta A_\mu]$ to the linear response, however the filled negative energy bands contribute total of $- \frac{1}{2} W_{nr}[\delta A_\mu]$, which leads to exact relation \eqref{exactHallC}.
To prove \eqref{DvsNRW} we first turn to the non-relativistic generating functional. In the leading order we have
\bea\nonumber
&&\!\!\!\!\!\!\!\!\!\!\!\!\!\!\!\!\!\!W_{\rm nr}[\delta A_\mu] =\frac{1}{4\pi} \int d^2x dt \Big[ Ad  A  \\
&&\!\!\!\!\!\!\!\!\!\!\!\!\!\!\!\!\!\!+\frac{1}{ \omega_c} |\vec \delta E|^2-\frac{2-\rg}{2m} \delta B^2 - \frac{3-\rg}{2} \ell^2  \delta B (\p_i \delta E_{i}) +\ldots\Big]\,.
\eea
In the LLL limit $m\rightarrow 0, \rg=2$ we find
\be\la{WNRprojected}
W^{m\rightarrow0}_{\rm nr}[\delta A_\mu] = \frac{1}{4\pi} \int AdA - \frac{1}{2} \ell^2 \delta B (\p_i \delta E_i) + \ldots\,.
\ee

\begin{figure*}
\centerline{\includegraphics[scale=0.8]{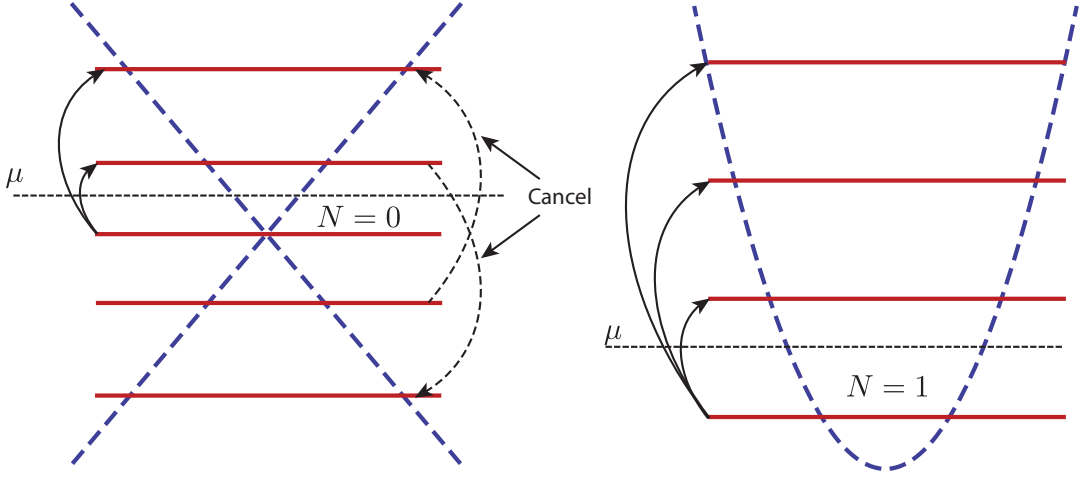}}
\caption{\textbf{Left,} Spectrum of Dirac operator in magnetic field. Dashed lines illustrate the transitions ``across the Fermi sea'', while solid lines illustrate the transitions between LLL and the excited levels. Total contribution to the linear response of the LLL of Dirac electrons from the transitions ``across the Fermi sea'' adds up precisely to $0$, leading to an exact relation \eqref{DvsNRW}.~\textbf{Right,} Spectrum of non-relativistic electrons in magnetic field. Transitions that contribute to the electromagnetic response of the LLL are qualitatively the same is in the Dirac case. The presence of the filled Dirac sea results in the overall factor of $1/2$ in the Hall conductivity of Dirac electrons.}
\label{diractransitions}
\end{figure*}

In fact, by dimensional analysis and the regularity of the massless limit only the terms linear in the electric field survive\footnote{This also can be seen, with some work, from the Eq.\eqref{Polanswer}}. These terms contribute to the (momentum dependent) Hall conductivity, which - in this limit - is can be calculated exactly \cite{can2014fractional}
\be\la{exactsigma}
\sigma^{\rm nr}_H(k) = \frac{1}{2\pi} \frac{1}{|k\ell|^2} s(k)\,,
\ee
where $s(k) = 1 - e^{-\frac{|k\ell|^2}{2}}$ is the static structure factor\cite{girvin1986magneto}. Eq.\eqref{exactsigma} agrees with the results produced from \eqref{Polanswer}.

The Dirac electrons are more tricky since in addition to the contribution of the LLL there are also, in principle, contributions from the transitions across the Dirac sea as shawn on Fig. \ref{diractransitions}. We have checked that these transitions sum up to zero in $|k\ell|^0,|k\ell|^2$, $|k\ell|^4$ and $|k \ell|^6$ orders of the momentum expansion. If we assume that these transitions do not contribute to {\it all} orders in the gradient expansion then the Hall conductivity is given by (using \eqref{eq:polDirac}-\eqref{GammaD})
\be
\sigma^{D}_H(k) = \frac{ e^{-\frac{|k\ell|^2}{2}}}{2\pi}\sum_{n=1}^\infty \frac{ 2^{-n-1} }{  n!}|k\ell|^{2 n-2} = \frac{\left(1 - e^{-\frac{|k\ell|^2}{2}}\right)}{4\pi |k\ell|^2} 
\ee
which leads leads to the exact relationship 
\be\la{exactHallC}
\sigma^{D}_H(k) = \frac{1}{2} \sigma^{\rm nr}_H(k)\,,
\ee
which is equivalent to \eqref{DvsNRW}.

There is, however, a subtlety as we have not explained how to take the massless limit of the Dirac generating functional \eqref{GenDirac}. Since the energy levels are given by $E_n =\pm v_F \sqrt{2\bar Bn}$ we should take the limit $v_F \rightarrow \infty$ which removes all of the energy levels except $E_0$. In this limit the term quadratic in the electric field indeed vanishes and the term linear in the electric field survives, however the term quadratic in magnetic field diverges linearly with $v_F$. In the non-relativistic case this term was removed by an appropriate choice of the $\rg$-factor $\rg=2$, but in the Dirac case there is no such mechanism. Thus, in order to ensure the regularity of the ``massless'' limit we must subtract this term ``by hand''. 

\section{Universality of the Large $N$ limit}
\label{sec:largeNap}
In this Section, we calculate the polarization tensor in the large $N$ limit and then re-derive it using the semiclassical approximation. The result of the semiclassical calculation agrees with previous work \cite{Simon1993}, but we will present a simpler method of calculation. Furthermore, we show that in the large $N$ limit, the result of the RPA calculation agrees with the Fermi liquid theory\cite{Pines:1966,Noieres:1964,Landau:1980}. 
\subsection{Large $N$ Limit of RPA}
\label{sec:largeN}
Working at large filling factors $N$ means that we consider a regime in which the density of electrons is much bigger than the external magnetic field
\begin{equation}
N=\frac{\bar \rho}{\bar B/2\pi} \gg 1.
\end{equation}
Non-interacting electrons in a weak magnetic field form a Fermi sphere. Furthermore, large filling also implies $k_F \ell \gg 1$. Therefore the gradient expansion in $k$ is valid in the range of momenta that satisfy $k\ell \sqrt{N}\sim 1$, which is the right regime for Landau's Fermi liquid theory.

First, we will explicitly take the large $N$ limit of the RPA result \eqref{Polanswer}. We will use the asymptotic form of Laguerre polynomial (valid in the leading order in $N$)
\begin{equation}
\lim_{N\to \infty} L_N^\alpha(x) \approx \frac{N^{\alpha/2}}{x^{\alpha/2}}e^{\frac{x}{2}}J_\alpha(2\sqrt{Nx})\,.
\end{equation}
In the expression \eqref{Polanswer} only the terms for $n\approx n^\prime$ contribute to the final result. This remains true for any $N$. Thus in the following we will use approximation
\begin{equation}
\frac{n^\prime!}{n!} \approx \frac{1}{N^{n-n^\prime}}\,,
\end{equation}
that is valid for $n\approx n^\prime \approx N$.

The generating function $\mathcal G^{as}(k,k^\prime,N)$ in the large $N$ limit takes form
\begin{widetext} 
	\begin{equation}
	\label{eq:genf1}
	\mathcal G^{\rm as}(k,k',N)=\sum_{n=1}^{\infty}\left(-\frac{\ell^2}{2}\right)^n\frac{ne^{\frac{\ell(|k^\prime|^2+|k|^2)}{4}}\left(\frac{(k \bar{k}')^n}{n-\omega }+\frac{(\bar{k} k')^n}{n+\omega }\right) J_n\left(2\sqrt{N\frac{k \bar{k}  \ell^2}{2}}\right) J_n\left(2\sqrt{N\frac{k' \bar{k}'  \ell^2}{2}}\right)}{(k \bar{k})^{n/2} (k' \bar{k}')^{n/2}},
	\end{equation}
\end{widetext}
where $\omega$ is the dimensionless frequency.

The asymptotic form of generating function $\mathcal G^{as}(k,k',N)$ agrees with the exact generating function $\mathcal G(k,k',N)$ up to sub-leading order in $N$, which can be checked order by order in the momentum expansion. We choose a frame where $\mathbf{k}=(k_1,0)$, in this case $k=\bar{k}=k_1$. It will be convenient to use the rescaled momentum $ q=k_1\ell\sqrt{2N}=k_1k_F\ell^2$.

Finally, using Eq.\eqref{eq:genf1} together with \eqref{Polanswer}, we obtain the polarization tensor for any $\rg$-factor
\bea
\label{eq:xx1}
\Pi^{11}(q,\omega)&=&-\frac{\bar \rho}{m}+\sum_{n=1}^{\infty}-\frac{n^4 \omega_c \left[J_n(q)\right]^2}{2\pi N(\omega^2-n^2)},\\\nonumber
\label{eq:yy1}
\Pi^{22}(q,\omega)&=&-\frac{\bar \rho}{m}+\sum_{n=1}^{\infty}-\frac{N n^2 \omega_c \left[J_{n-1}(q)-J_{n+1}(q)\right]^2}{2\pi (\omega^2-n^2)}\\
&&-\rg\sum_{n=1}^{\infty}\frac{ n^2 \omega_c q \left[J_{n-1}(q)-J_{n+1}(q)\right]J_{n}(q)}{4\pi (\omega^2-n^2)}\nonumber\\
&&-\rg^2\sum_{n=1}^{\infty}\frac{q^2  n^2 \omega_c \left[J_n(q)\right]^2}{32 N\pi (\omega^2-n^2)},\\
\nonumber
\Pi^{12}(q,\omega)&=&-\sum_{n=1}^{\infty}\frac{i N n^2 \omega\omega_c J_n(q)q\left[ J_{n-1}(q)-J_{n+1}(q)\right]}{\pi(\omega^2-n^2)}\\ \label{eq:xy1}
&&-\rg\sum_{n=1}^{\infty}\frac{i\omega n^2 \omega_c \left[J_n(q)\right]^2}{4\pi (\omega^2-n^2)}.
\eea
Note that using Eqs.\eqref{Pi00g}-\eqref{Piijg} one can restore all of the components of the polarization tensor at vanishing $\rg$-factor.

Remarkably, the infinite sums for each component of the non-relativistic polarization tensor can be evaluated in a closed form. The details of the calculation are presented in the Appendix \ref{sec:largeN-NRsum}. The results are written for the conductivity tensor \eqref{sigmaToPi}. 
\bea
\label{eq:largeNxx}
&&\sigma^{11}(q,\omega)=\frac{i N}{\pi}\left(-\frac{\omega}{q^2}+\frac{\pi \omega^2 J_{\omega}(q)J_{-\omega}(q)}{q^2\sin(\pi\omega)}\right),\\ \nonumber
&&\sigma^{22}(q,\omega)=\frac{i N}{\pi}\left(-\frac{\omega}{q^2}+\frac{\pi \omega^2 J_{\omega}(q)J_{-\omega}(q)}{q^2\sin(\pi\omega)} \right.\\ \nonumber
&&\left.+\frac{\pi J_{1+\omega}(q)J_{1-\omega}(q)}{ \sin(\pi\omega)}\right)+\frac{i \rg  q}{8 \sin (\pi  \omega )}\frac{\partial}{\partial q}\left[J_{\omega}(q)J_{-\omega}(q)\right]\\ \label{eq:largeNyy}
&&-\frac{i\rg^2  q^2}{64 \pi  N \omega}\left(1-\frac{\pi \omega}{\sin(\pi\omega)}J_{\omega}(q)J_{-\omega}(q)\right)\\\nonumber
&&\sigma^{12}(q,\omega)=-N\frac{ \omega}{2q} \frac{1}{\sin(\pi \omega)}\frac{\partial}{\partial q} \big(J_{\omega}(q)J_{-\omega}(q)\big)\\ \label{eq:largeNxy}
&&+\frac{\rg}{8 \pi } \left(1-\frac{\pi \omega}{\sin(\pi\omega)}J_{\omega}(q)J_{-\omega}(q)\right)\,.
\eea
Eqs.\eqref{eq:largeNxx} - \eqref{eq:largeNxy} are the main result of this Section. Next we will compare these results to a semiclassical computation.

\subsection{Semiclassical Computation}
\label{sec:semi}
\subsubsection{Review of the Fermi liquid theory with a $\rg$-factor}	
\label{sec:Boltz}
In this Section, we review the derivation of Boltzmann's equation mostly to fix the notation. The derivation follows closely to the bosonization of Fermi liquid \cite{CastroNeto:1994,Haldane:1994,Polchinski:1992ed,Shankar:1993pf}.
We assume a system of two dimensional non-interacting spinless fermions with Fermi momentum $k_F$ with mass $m$ in  magnetic field $B(x,t)= \bar B + b(x,t)$ and electric field $\vec E(x,t)$. We will assume that $b(x,t)$ and $\vec E(x,t)$ are weak and slowly varying. We denote the distribution function as $f(\mathbf{K},\mathbf{x},t)$. The collective modes are described by the perturbation of distribution function 
\begin{equation}
f(\mathbf{K},\mathbf{x},t)=f^0(\mathbf{K}) +\delta f(\mathbf{K},\mathbf{x},t). 
\end{equation} 
The perturbations of the distribution function caused by weak fields are also assumed to be weak.
Where the unperturbed distribution function is 
\begin{equation}
f^0(\mathbf{K})=\Theta(k_F-|\mathbf{K}|),
\end{equation}	
where $\Theta(x)$ is the step function. Employing the Boltzmann equation\cite{Ying:1968,Platzman:1967,Wen:1995}, we obtain the time evolution equation for the distribution function $ f(\mathbf{K},\mathbf{x},t)$. 
\bea\nonumber
&&\!\!\!\!\!\!\partial_tf(\mathbf{K},\mathbf{x},t)+ \vec{v}(\mathbf{K})\cdot\vec{\nabla}_\mathbf{x} f(\mathbf{K},\mathbf{x},t)\\ \nonumber&&\!\!\!\!\!\!+\big(\vec{E}+\frac{\rg}{4m}\vec{\nabla}_\mathbf{x} B(\mathbf{x},t)+\vec{v}(\mathbf{K})\times \vec{B}(\mathbf{x},t)\big)\cdot\vec{\nabla}_\mathbf{K} f(\mathbf{K},\mathbf{x},t)=0,
\eea
 where $\vec{v}(\mathbf{K})=\vec{\nabla}_\mathbf{K}\epsilon_\mathbf{K}$ is the group velocity and $\epsilon_\mathbf{K}$ is the non-relativistic dispersion relation. We also introduce a vector, normal to the Fermi surface via $\vec v(\mathbf{K}) = v_F \vec{n}_\theta$. Note that we included the term $\frac{\rg}{4m}\vec{\nabla}_\mathbf{x} B(\mathbf{x},t)\vec \nabla_{\mathbf K}f(\mathbf{K},\mathbf{x},t)$ which is necessary to account for the finite $\mathrm g$-factor.
 
\begin{figure}
\centerline{\includegraphics[scale=0.5]{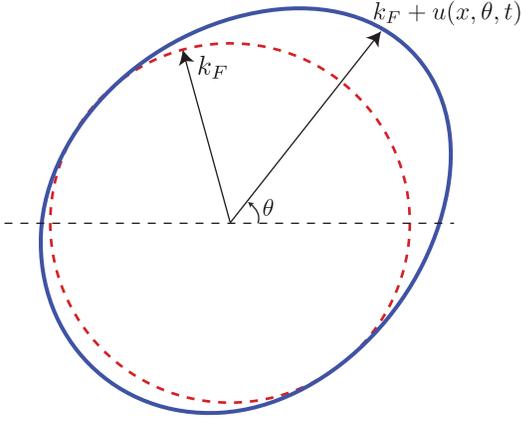}}
\caption{Fluctuating Fermi surface. The function $u(x,\theta, t)$ describes the fluctuations of the surface in space and time. The harmonics $u_n(x,t)$ describe the dipolar, quadruploar, etc. deformations of the Fermi surface. }
\label{Fermi}
\end{figure}

 In the low energy limit we take the momentum to be $|\mathbf{K}| = k_f + u(x,\theta,t)$ (see Fig. \ref{Fermi}). Then the perturbations of the distribution function occur only close the the Fermi surface
\begin{equation}
\delta f(\mathbf{K},\mathbf{x},t)=u(\theta,\mathbf{x},t)\delta(k_F-|\mathbf{K}|),
\end{equation}  
where $\theta$ is the direction of $\mathbf{K}$ on the Fermi surface.

Then Boltzmann equation takes form\cite{GNRS1,GNRS2}
\bea\nonumber
&&\partial_t u(\theta,\mathbf{x},t)+ v_F \vec{n}_\theta\cdot\vec{\nabla}_\mathbf{x}u(\theta,\mathbf{x},t)  -\omega_c\partial_\theta u(\theta,\mathbf{x},t)\\\label{eq:boltz1}
&&-\vec{n}_\theta\cdot\left(\vec{E}(\mathbf{x},t)+\frac{\rg}{4m}\vec{\nabla}_\mathbf{x} B(\mathbf{x},t)\right)=0,
\eea
where $v_F=\frac{k_F}{m}$ is the Fermi velocity, and $\vec{n}_\theta$ is the normal vector to the Fermi surface. We ignore terms that are second order in $\vec{E}(\mathbf{x},t)$, $\vec{b}(\mathbf{x},t)$ and $\delta f(\mathbf{K},\mathbf{x},t)$. The charge density of the electrons can be written in terms of $u(\theta,\mathbf{x},t)$ as follows
\begin{equation}
\label{eq:rhoboltz}
\rho(\mathbf{x},t)=\int \frac{d^2 \mathbf{K}}{(2\pi)^2}f(\mathbf{K},\mathbf{x},t)=\bar \rho+\int d\theta\frac{k_F}{(2\pi)^2}  u(\theta,\mathbf{x},t),
\end{equation}	
where the background charge density is given by
\begin{equation}
\bar \rho=\int \frac{d^2 \mathbf{k}}{(2\pi)^2}f^0(\mathbf{K})=\frac{k_F^2}{4\pi}\,.
\end{equation}

At nonzero $\mathrm g$-factor the current density is defined as (cf. Eq. \eqref{current-g})
\be\nonumber
J^i(\mathbf{x},t)=\int \frac{d^2 \mathbf{K}}{(2\pi)^2}f(\mathbf{K},\mathbf{x},t)v^i(\mathbf{K})+\frac{\rg}{4m}\epsilon^{ij}\partial_j \rho\,.\\
\ee
Which in terms of $u(\theta, \mathbf x, t)$ is given by
\bea\nonumber
J^i(\mathbf x,t)&=& \frac{k_F v_F}{2\pi}\int \frac{d\theta}{2\pi} n^i_\theta u(\theta,\mathbf{x},t)\\
\label{eq:jiboltz}
&&+\frac{\rg}{4m}\frac{k_F}{2\pi} \epsilon^{ij}\partial_j \int \frac{d\theta}{2\pi} u(\theta,\mathbf{x},t).
\eea
The equation (\ref{eq:boltz1}),(\ref{eq:rhoboltz}) and (\ref{eq:jiboltz}) are the key ingredients for the semiclassical calculations.

\subsubsection{Semiclassical calculation for the non-relativistic polarization tensor }
\label{sec:semi1}

 We will work in the temporal gauge $A_0=0$. In this gauge the electric field is given by $E_i(\vec{q},\omega)=i\omega A_i(\vec{q},\omega)$. We decompose $u(\theta,\mathbf{x},t)$ into Fourier modes
\begin{equation}
u(\theta,\mathbf{x},t)=\int \frac{d\Omega d^2\mathbf{k}}{(2\pi)^3}\sum_{n=-\infty}^{\infty}u_n(\mathbf{k},\Omega)e^{i n \theta}e^{i\mathbf k\cdot  \mathbf x}e^{-i\Omega t}\,\,.
\end{equation}	
Next, we will fix the frame where $\mathbf{k}=(k_1,0)$ and introduce the notation 
\be
q=k_1\ell\sqrt{2N}=k_1k_F\ell^2\,.
\ee
Then Boltzmann equation (\ref{eq:boltz1}) takes form
\bea\nonumber
	(\omega&+&n)u_n(q,\omega)-\frac{q}{2}\left(u_{n+1}(q,\omega)+ u_{n-1}(q,\omega)\right)\\\nonumber
	&&+\omega\left[\delta_{n,1}\left(A_z+\rg\frac{q^2}{32N\omega}(A_z- A_{\bar z})\right)\right.+\\\label{eq:boltz2}
	&&\left.\delta_{n,-1}\left(A_{\bar{z}}+\rg\frac{q^2}{32N\omega}(A_z - A_{\bar z})\right)\right]=0.
\eea

The solution of above equation of motions for $u_n(q,\omega)$ with $n>1$ and $n<-1$ is 
\begin{align}
\label{eq:sol1}
u_n(q,\omega)=F(q,\omega)J_{n+\omega}(q) \qquad (n>0)\\
\label{eq:sol2}
u_n(q,\omega)=(-1)^n G(q,\omega)J_{-n-\omega}(q) \qquad (n<0)
\end{align}
where $J_\nu(x)$ is the Bessel function of the first kind. The functions $F(q,\omega)$ and $G(q,\omega)$ depend on the external  field and are not fixed by the equations for $|n|>1$. We will fix these functions using the equations of motion for $u_{-1},u_0,u_1$.

The equation of motion for $u_0(q,\omega)$ gives us 
\begin{equation}
\label{eq:sol3}
u_0(q,\omega)=\frac{q}{2}\left(F(q,\omega)J_{1+\omega}(q)- G(q,\omega)J_{1-\omega}(q) \right).
\end{equation}
Using (\ref{eq:sol1}),(\ref{eq:sol2}) and (\ref{eq:sol3}) in the equation of motion for $u_1(q,\omega)$ and $u_{-1}(q,\omega)$ we find 
\bea\nonumber
&&\!\!\!\!\!F(q,\omega)=\frac{\pi \omega}{\sin(\pi\omega)}\left(J_{-1-\omega}(q)\left(A_z+\rg\frac{q^2}{32N\omega}(A_z- A_{\bar z})\right)\right.\\\label{eq:F}
&&\!\!\!\!\!\left.+J_{1-\omega}(q)\left(A_{\bar{z}}+\rg\frac{q^2}{32N\omega}(A_z - A_{\bar z})\right)\right),\\
\nonumber
&&\!\!\!\!\!G(q,\omega)=-\frac{\pi \omega}{\sin(\pi\omega)}\left(J_{-1+\omega}(q)\left(A_z+\rg\frac{q^2}{32N\omega}(A_z- A_{\bar z})\right)\right.\\\label{eq:G}
&&\!\!\!\!\!\left.+J_{1+\omega}(q)\left(A_{\bar{z}}+\rg\frac{q^2}{32N\omega}(A_z - A_{\bar z})\right)\right),
\eea
where we used the following Bessel function identity
\begin{equation}
\label{eq:bes1}
J_{1-\omega}(q)J_{1+\omega}(q)-J_{-1-\omega}(q)J_{-1+\omega}(q)=\frac{4\omega\sin(\pi\omega)}{\pi q^2}.
\end{equation}
Functions $u_1(q,\omega)$ and $u_{-1}(q,\omega)$ are then given by
\be\la{u1}
u_1(q,\omega)=F(q,\omega)J_{1+\omega}(q),\quad u_{-1}(q,\omega)=-G(q,\omega)J_{1-\omega}(q)\,.
\ee
To calculate the response functions in terms of the applied electric field, we write equation (\ref{eq:jiboltz}) in terms of the Fourier modes
\begin{equation}
J^1=\frac{N\omega_c}{2\pi}(u_1+u_{-1}),\quad
J^2=\frac{iN\omega_c}{2\pi}(u_{1}-u_{-1})-\frac{i \rg q \omega_c}{8\pi}u_0,
\end{equation} 
where $N=\frac{1}{2} k_F^2 \ell^2$ is the number of filled Landau levels.

Using Eqs. \eqref{u1} we can derive the current density in terms of vector potential in the usual form
\begin{equation}
\label{eq:PiLade}
J^i(q,\omega)=\sigma^{ij}(q,\omega)E_j(q,\omega)\,,
\end{equation}
from where we can extract the polarization tensor which is again given exactly by \eqref{eq:largeNxx}-\eqref{eq:largeNyy} combined with Eqs. \eqref{Pi00g}-\eqref{Piijg}. Reducing Eq.\eqref{eq:PiLade} to the form of Eqs. \eqref{eq:largeNxx}-\eqref{eq:largeNyy} involves non-trivial manipulations with the Bessel functions. We leave these details to the Appendix \ref{sec:largeN-NRsum}. We conclude that the RPA approximation in the large $N$ limit is equivalent to the semiclassical approximation for {\it any} value of the $\mathrm g$-factor.


\section{Conclusion}
We have calculated the electromagnetic response of IQH state of non-relativistic and massless Dirac electrons to {\it all} orders in the gradient expansion. In the non-relativistic case we obtained a simple closed form expression \eqref{Polanswer} wich agrees with the one loop calculation from the previous work \cite{lopez1991fractional} for non-relativistic electrons as well as general (non-linear) structure of the generating functional \cite{klevtsov2009bergman, klevtsov2015quantum}. The method we used is extended naturally to the massless Dirac theory in magnetic field. We explicitly check that  the polarization tensors of non-relativistic and Dirac electrons match in the large $N$ limit up to substitution $N\rightarrow N+1/2$.
The extra $1/2$ is due to the Berry phase of the Dirac cone. Furthermore, in the Dirac case, we check that the $\Omega^2$ and $k^2$ corrections to the Hall conductivity satisfy the extra relation \eqref{Dirack2} imposed by the Lorentz invariance\cite{GRS}.

We have used the semiclassical approximation to calculate the electromagnetic response function of the Fermi liquid in weak constant background magnetic field, the polarization tensor can be written in a closed form, given in terms of Bessel's functions, and agrees with the previous work \cite{Simon1993}, however we have used a simpler method of calculation. Our computation can be easily modified to include the effect of short range interactions via introducing the Landau parameters \cite{GNRS2}.  The results, which includes short range interaction, can still be derived in a closed form. Next, we showed explicitly, that the large $N$ limit of RPA calculation in the non-relativistic case matches the semi-classical approximation at the leading and sub-leading order in $N$, without including the short range interactions. The agreement implies the equivalence of Fermi liquid theory in a weak background magnetic field and large $N$ limit of RPA calculation. Finally, in view of the previous result we see that the Fermi liquid theory must be modified by the Berry phase effects in order to work for the Dirac fermions. This effect can be easily incorporated via the substitution $N \rightarrow N+1/2$.

We expect that our computations will find many applications to quantum Hall physics. The explicit expression for the polarization tensor is necessary in composite fermion \cite{lopez1991fractional} and boson \cite{zhang1989effective} approaches to fractional quantum Hall (FQH) states. These results can also serve as a starting point to accounting for lattice, quenched disorder and weak interactions corrections to the linear response theory. Moreover, some of the gradient corrections to the transport coefficients, under certain symmetry assumptions, cary universal information about the quantum Hall states\cite{hoyos2012hall, bradlyn2012kubo, golkar2014effective, gromov2014density, can2014fractional}, thus the knowledge of these corrections as well as general method of their computation is of its own interest. The large $N$ results should be useful in the recently proposed theory of composite fermions\cite{Son:2015xqa}, where the latter are viewed as neutral Dirac fermions interacting with an internal gauge field. Finally, all of the exact results are useful in testing the recently discovered set of dualities in $2+1D$ \cite{metlitski2015particle, seiberg2016duality, karch2016particle}.

The methods used in the present paper are suitable for the calculation of the gravitational (or viscoelastic) and mixed electromagnetic-gravitational response functions of quantum Hall fluids in curved space\cite{wen1992shift, gromov2015framing} as well as more general Newton-Cartan\cite{son2013newton, gromov2015thermal, bradlyn2015low} backgrounds. We will present the detailed computations of these responses in a separate publication.

{\it Acknowledgments.} It is a pleasure to thank Alexander G. Abanov and Dam T. Son for stimulating discussions and useful suggestions. 

A.G. was supported by the Kadanoff Fellowship. D.X.N. is supported by the Chicago MRSEC, which is
funded by NSF through grant DMR-1420709.

\appendix
\section{Generating functional summary}
For the reader's convenience we list together all of the final expressions derived in the Section II in terms of dimensionless momentum $q= \frac{kl}{\sqrt{2}}$ and for arbitrary $\mathrm g$-factor.
\be
\delta S_{\rg} = \frac{\rg}{4m}\int dtd^2x  B \psi^\dag \psi\,.
\ee

The generating function is given by
\begin{widetext}
\bea
 \!\!\!\!\!\!\!\mathcal G(q,q^\prime;N)  =\sum_{n\geq N,n^\prime< N}\left(-1\right)^{n-n^\prime}\frac{n^\prime!}{n!}\left(\frac{(q\bar{q}^\prime)^{n-n^\prime}}{n-n^\prime-\omega}+\frac{(\bar{q}q^\prime)^{n-n^\prime}}{n-n^\prime+\omega}\right) L_{n^\prime}^{n-n^\prime}\left(\frac{|q|^2}{2}\right)L_{n^\prime}^{n-n^\prime}\left(\frac{|q^\prime|^2}{2}\right)\,,
\eea
\end{widetext}

\newpage

The vertices are given by the following relations

\bea
\hat{\mathcal P}^0_{\rm nr}(q) &=&1\,,\\
\hat{\mathcal P}^{\bar z}_{\rm nr}(q) &=&-\frac{1}{2\sqrt{2}m\ell}\left(2\p_{\bar{q}} - \left(1- \frac{\rg}{2}\right)q\right)\,,\\
\eea
where we have also added the dependence on the $g$-factor that describes the non-minimal coupling of the electrons to the magnetic field due to the intrinsic magnetic moment.
The polarization operator is given by
\be\la{W-final}
\Pi_{\mu\nu} = \frac{m}{4\pi}e^{-|q|^2}\lim_{q\rightarrow -q^\prime}\hat{\mathcal P}_\mu(q)\hat{\mathcal P}_\nu(q^\prime)G(q,q^\prime;N) + \Pi_c^{\mu\nu}
\ee
and
\be
W^{(2)}_{{\mbox \rm c}} = \frac{N\omega_c}{4\pi} \int \frac{d^2 \vec{q}}{(2\pi)^2} |\delta A(\vec q, 0)|^2
\ee

\section{Derivation of \eqref{diagramanswer}}
\la{Derivation}
\subsubsection{Summation over $b$ subspace}

The first step in evaluation of \eqref{trace} is to perform the summation over the Fock space generated by $b, b^\dag$ operators. This can be done easily because the $b, b^\dag$ operators completely factorize from the expression for the vertices \eqref{OMG-0}-\eqref{OMG-2}, because the perturbed action does not depend on $b$ and $(a, a^\dag)$ commute with $(b, b^\dag)$. We compute the trace over the Fock spaces (suppressing the frequency integration)
\begin{widetext}
\bea\nonumber
&&\Tr_{a,b} G_0V^{(1)} G_0 V^{(1)}= \sum_{n,n^\prime,m,m^\prime} \langle nm |G_0|nm\rangle\langle n m|V^{(1)} |n^\prime m^\prime\rangle\langle n^\prime m^\prime|G_0 |n^\prime m^\prime \rangle\langle n^\prime m^\prime|V^{(1)}| nm\rangle \\
&=&\sum_{n,n^\prime,m,m^\prime}  \frac{1}{ \omega - E_{n}}\frac{1}{ \omega^\prime - E_{n^\prime}}\langle n m|V^{(1)} |n^\prime m^\prime\rangle\langle n^\prime m^\prime|V^{(1)}| nm\rangle
\eea
The matrix elements $\langle n^\prime m^\prime|V^{(1)}| nm\rangle$ factorize as
\be\la{factor}
\langle n^\prime m^\prime|V^{(1)}| nm\rangle= \langle m^\prime|e^{-\frac{k\ell}{\sqrt{2}}b}e^{\frac{\bar{k}\ell}{\sqrt{2}}b^\dag}| m\rangle\Big|_b\cdot\langle n^\prime |\mathcal V^{(1)}| n\rangle\Big|_a
\ee
because $a$ commutes with $b$. In Eq. (\ref{factor}) $\langle m | X | m^\prime\rangle\Big|_b$ means that the average value of operator $X$ is computed in the Fock space generated by the $b^\dag$. Then
\bea\nonumber
Tr_{a,b} G_0V^{(1)}G_0 V^{(1)} =&&\sum_{n,n^\prime,m,m^\prime,k,q}\frac{1}{ \omega - E_{n}}\frac{1}{ \omega^\prime - E_{n^\prime}}\langle m|e^{-\frac{k\ell}{\sqrt{2}}b}e^{\frac{\bar{k}\ell}{\sqrt{2}}b^\dag}| m^\prime\rangle\Big|_b\\ \nonumber
&\times& \langle m^\prime|e^{-\frac{q\ell}{\sqrt{2}}b}e^{\frac{\bar{q}\ell}{\sqrt{2}}b^\dag}| m\rangle\Big|_b \langle n^\prime | V_\mu^{(1)}| n\rangle\Big|_a
\langle n^\prime | V_\nu^{(1)}| n\rangle\Big|_a \delta A_\mu(k) \delta A_\nu(q) \\ \nonumber
&=&\sum_{n,n^\prime,m,k,q}\frac{1}{ \omega - E_{n}}\frac{1}{ \omega^\prime - E_{n^\prime}}\langle m|e^{-\frac{k\ell}{\sqrt{2}}b}e^{\frac{\bar{k}\ell}{\sqrt{2}}b^\dag}e^{-\frac{q\ell}{\sqrt{2}}b}e^{\frac{\bar{q}\ell}{\sqrt{2}}b^\dag}| m\rangle\Big|_b\\ 
&\times& \langle n^\prime | V_\mu^{(1)}| n\rangle\Big|_a
\langle n^\prime | V_\nu^{(1)}| n\rangle\Big|_a\delta A_\mu(k)\delta A_\nu(q)\, ,
\eea
where in the last line we have used that $|m\rangle$ form a complete basis in the Fock space generated by $b$.
\be
\sum_{m^\prime}|m^\prime\rangle\langle m^\prime| = \mathbf 1_b\,, 
\ee
where $\mathbf 1_b$ is an identity operator in the Fock space spanned by $b$ operators. We have, thus, established that in all of the components of the generalized polarization operator the summation over $m$ can be done explicitly and amounts to the computation of the sum
\bea
\sum_{m} \langle m|e^{-\frac{k\ell}{\sqrt{2}}b}e^{\frac{\bar{k}\ell}{\sqrt{2}}b^\dag}e^{-\frac{q\ell}{\sqrt{2}}b}e^{\frac{\bar{q}\ell}{\sqrt{2}}b^\dag}| m\rangle &&= \frac{1}{\pi}\int d\alpha e^{|\alpha|^2} \langle 0 | e^{\alpha b} e^{-\frac{k\ell}{\sqrt{2}}b}e^{\frac{\bar{k}\ell}{\sqrt{2}}b^\dag}e^{-\frac{q\ell}{\sqrt{2}}b}e^{\frac{\bar{q}\ell}{\sqrt{2}}b^\dag}  e^{\bar{\alpha} b^\dag}|0\rangle\\
&&= \frac{2\pi}{\ell^2} e^{-\frac{|k\ell|^2}{2}}\delta^{(2)}(k+q)\,.
\eea
In the first line we replaced the summation in $m$ with integration over the coherent states (we explain how to do it in the Appendix \ref{cohstates}). In resume: for any component of the polarization tensor summation over $m$ can be replaced by $\frac{2\pi}{\ell^2} e^{-\frac{|k\ell|^2}{2}}\delta^{(2)}( k+ q)$. This delta function is the manifestation of the momentum conservation - after $b$-summation the fully filled Landau level looks translationally invariant.

\subsubsection{Frequency integral}
Next we perform the trace over time and frequency
\bea\nonumber
&&\Tr_t G_0V^{(1)} G_0 V^{(1)} = \sum_{t}\langle t | G_0V^{(1)}G_0V^{(1)} |t\rangle \\ \nonumber
&=& \sum_{t,\omega}  \sum_{t^\prime,\omega^\prime}\langle t|\omega\rangle \langle \omega | G_0 | \omega\rangle \langle \omega | t^\prime \rangle \langle t^\prime | V^{(1)}_{n n^{\prime}}  |t^\prime\rangle \langle t^\prime | \omega^\prime \rangle\langle \omega^\prime | G_0 | \omega^\prime\rangle \langle \omega^\prime |t\rangle \langle t| V^{(1)}_{n^\prime n} |t \rangle \\ \nonumber
&=&  \sum_{n,n^\prime}\sum_{t,\omega}  \sum_{t^\prime,\omega^\prime}e^{it(\omega - \omega^\prime)}e^{-it^\prime(\omega - \omega^\prime)} \frac{1}{ \omega - E_n} V_{n n^\prime}^{(1)} (t) \frac{1}{ \omega^\prime - E_{n^\prime}} V_{n^\prime n}^{(1)} (t^\prime)
 \\ \nonumber
&=&   \sum_{n,n^\prime}\sum_{t,\omega, \Omega}  \sum_{t^\prime,\omega^\prime,\Omega} e^{it(\omega - \omega^\prime - \Omega)}e^{-it^\prime(\omega - \omega^\prime-\Omega^\prime)}\frac{1}{ \omega - E_n} V_{n n^\prime}^{(1)} (\Omega) \frac{1}{ \omega^\prime - E_{n^\prime}} V_{n^\prime n}^{(1)} (\Omega^\prime) \\ \nonumber
&=& \sum_{n,n^\prime}\sum_{\omega, \Omega}  \sum_{\omega^\prime,\Omega} \delta(\omega - \omega^\prime - \Omega)\delta(\omega - \omega^\prime-\Omega^\prime)\frac{1}{ \omega - E_n} V_{n n^\prime}^{(1)} (\Omega) \frac{1}{ \omega^\prime - E_{n^\prime}} V_{ n^\prime n}^{(1)} (\Omega^\prime) \\ \nonumber
&=&\sum_{n,n^\prime}\sum_{\omega, \Omega} \frac{1}{ (\omega + \Omega) - E_n} V_{n n^\prime}^{(1)} (\Omega) \frac{1}{ \omega - E_{n^\prime}} V_{ n^\prime n}^{(1)} (-\Omega) \\ \nonumber
&=& \sum_{n,n^\prime} \int \frac{d\Omega}{2\pi}\frac{d\omega}{2\pi}\frac{1}{ (\omega + \Omega) - E_n} \frac{1}{ \omega - E_{n^\prime}} V_{n n^\prime}^{(1)} (\Omega) V_{n^\prime n}^{(1)} (-\Omega)\,,
\eea
where we have introduced a shorthand $V_{nn^\prime}$ for matrix elements $\langle n|V|n^\prime\rangle$.
To perform the frequency integration we rewrite the fraction as a sum
\be
\frac{1}{ (\omega + \Omega) - E_n} \frac{1}{ \omega - E_{n^\prime}}  = \left(\frac{1}{ (\omega + \Omega) - E_n}  -\frac{1}{ \omega - E_{n^\prime}} \right) \frac{-1}{\Omega - (E_{n}-E_n^\prime)}\,.
\ee
and take only first $N$ poles in the integral over $\omega$. This integration will project onto Hilbert space of the first $N$ Landau levels. When this is done we have
\bea\nonumber
&&Tr_t G_0V^{(1)} G_0 V^{(1)}=\int \frac{d\Omega }{2\pi}\left(\sum_{n,n^\prime} \frac{\theta (N-n) V_{n^\prime n}^{(1)} (\Omega) V_{n n^\prime }^{(1)} (-\Omega)}{E_{n^\prime} - E_n - \Omega} - \frac{\theta (N-n^\prime)V_{n^\prime n}^{(1)} (-\Omega) V_{n n^\prime }^{(1)} (\Omega)}{E_{n^\prime} - E_n - \Omega} \right) \\ \nonumber
&=& \int \frac{d\Omega }{2\pi}\sum_{n\leq N, n^\prime >N} \left(\frac{V^{(1)}_{n n^\prime} (\Omega) V^{(1)}_{n^\prime n} (-\Omega)}{E_{n^\prime} - E_n - \Omega} + \frac{V_{n^\prime n}^{(1)}(\Omega) V_{n n^\prime}^{(1)} (-\Omega)}{E_n - E_{n^\prime}+ \Omega }\right)\,.
\eea
This is the final outcome of the computation (we have suppressed the dependence on the momentum). This computation yields \eqref{diagramanswer}. 
\end{widetext}

\section{Coherent states}
\la{cohstates}

In this Appendix we will describe the coherent states that will be useful for multiple calculations.  Here we follow Perelomov \cite{perelomov1977generalized}, but customize the notations to agree with the main text. 

\subsection{Heisenberg-Weyl group}

We define Heisenberg-Weyl algebra via relations
\be
[a,a^\dag] = 1\,, \qquad [a,1]=[a^\dag,1]=0\,,
\ee
an arbitrary element of the algebra is given by a linear combination
\be
W = is\cdot {\bf{1}} + qa^\dag-\bar{q}a,
\ee
where $s$ is real and $q$ is complex.

We want to exponentiate the algebra into the group. Arbitrary Heisenbrg-Weyl group element is given by
\bea\nonumber
e^W &&= e^{is} \cdot e^{qa^\dag - \bar q a} = e^{is}  e^{qa^\dag}e^{-\bar q a}e^{-\frac{1}{2}[qa^\dag,-\bar qa]}\\
&&=e^{is} e^{-\frac{|q|^2}{2}}  e^{qa^\dag}e^{-\bar q a},
\eea
where we have used $e^{A+B} = e^{-\frac{1}{2}[A,B]}e^Ae^B$, which is true for linear combinations of creation/annihilation operators. We also denote 
\be
D(q) =e^{qa^\dag - \bar q a}\,.
\ee
These operators form a representation of the Heisenberg-Weyl group. Representations for different values of $s$ are inequivalent. For fixed value of $s$ all representations are unitary equivalent. So from now on we fix $s$ and drop $e^{is}$ factor.

We can freely switch between $D(q)$ and $e^{qa^\dag}e^{-\bar q a}$ at the cost of an exponent, that is
\be
D(q) = e^{-\frac{|q|^2}{2}}  e^{qa^\dag}e^{-\bar q a}\,.
\ee
Operators $D(q)$ have the following multiplication rule
\be
D(q)D(k) = e^{i{\mbox{Im}}(q\bar k)} D(k+q)\,.
\ee
This can be checked using the following simple identities
\bea\label{perm}
e^{ca}f(a^\dag) &=&  f(a^\dag + c) e^{ca}\,,\\
e^{ca^\dag} f(a) &=& f(a-c) e^{ca^\dag}\,.
\eea
These relations can be used to prove the multiplication law. The latter can be obviously generalized as follows
\bea
\prod_{i=M}^{i=1}D(q_i)  = e^{i\sum_{i<j}{\mbox{Im}}(q_j\bar q_i)} D\left(\sum_{i=1}^{M}q_i\right)\,.
\eea

The multiplication law implies the permutation relations
\be
D(q)D(k) = e^{2i{\mbox{Im}}(q\bar k)} D(k)D(q)\,.
\ee

\subsection{Generalized coherent states}

Operators $a$, $a^\dag$ naturally generate a Fock space $\mathcal{H}$ with an orthnormal basis
\be
|n\rangle = \frac{a^\dag}{\sqrt{n!}}|0\rangle\,,
\ee
where $|0\rangle$ is defined via $a|0\rangle=0$. Consider an arbitrary state $|\Psi_0\rangle \in \mathcal{H}$. States of the form
\be
D(q) |\Psi_0\rangle = |q\rangle
\ee
are generalized coherent states. One gets usual coherent states choosing $|\Psi_0\rangle = |0\rangle$. Most of relations for coherent sates hold for any $|\Psi_0\rangle$. The overlap of the coherent sates is
\be
\langle q|k\rangle = e^{i{\mbox{Im}}(k\bar q)}\langle\Psi_0 | D(k-q)|\Psi_0\rangle \qquad |\langle q|k\rangle|^2 \equiv \rho(k-q)
\ee 
Also we have
\be
D(k)|q\rangle =  e^{i{\mbox{Im}}(k\bar q)} |k+q\rangle\,.
\ee
Since the Fock space $\mathcal{H}$ is projected $D(k)$ acts on the $q$-plane by translations. Therefore an invariant (under the action of Heisenberg-Weyl group) measure is
\be\label{measure}
d\mu(k) = C dk_1dk_2\,, \qquad {\mbox{with}} \qquad k = k_1+ik_2\,,
\ee
where $C$ is arbitrary constant to be fixed momentarily.
Consider an operator
\be
A = \int d\mu(k) |k\rangle\langle k|\,.
\ee
We find that for any $k$ we have $[D(k),A] = 0$, thus $A=\lambda \hat{1}$ due to Schur's lemma. We also can always choose $C$ to set $\lambda =1$. We take $C=\frac{1}{\pi}$ then resolution of identity takes form (this particular value of $C$ will be explained shortly)
\be\label{unity}
 \int \frac{dk_1dk_2}{\pi} |k\rangle\langle k| = \hat{1}\,.
\ee
We now present some relations that are valid {\it only} for $|\Psi_0\rangle = |0\rangle$.
\be
D^\dag(q) a D(q) = a + q\,.
\ee
and
\be
a|k\rangle = k|k\rangle\,.
\ee

Similarly we have
\be
|k\rangle=D(k)|0\rangle = e^{-\frac{|k|^2}{2}}e^{ka^\dag}|0\rangle = \sum_{n=0}^\infty \frac{k^n}{\sqrt{n!}} |n\rangle
\ee
Using the last relation we find
\be
|\langle k|0\rangle|^2 = \rho(k) = e^{-|k|^2} \qquad |\langle k|q\rangle|^2 = \rho(q-k) = e^{-|k-q|^2} 
\ee
Noticing that (\ref{unity}) is equivalent to $\int d\mu(k) \rho(k) = 1$ we find that $C=\frac{1}{\pi}$ as advertised.

We want to be able to evaluate traces in $\mathcal{H}$. In the Fock basis we have
\bea\nonumber
\Tr O& =& \sum_n \langle n| O |n\rangle =\sum_{n} \int d\mu(k)d\mu(q) \langle q|n\rangle\langle n| k\rangle\langle k| O |q\rangle \\
&=& \int d\mu(k) d\mu(q) \langle q|k\rangle \langle k| O |q\rangle
\eea
Using resolution of identity
\be
\Tr O =  \int d\mu(q) \langle q| O |q\rangle = \int d\mu(q) e^{-|q|^2}\langle0|e^{\bar qa}Oe^{qa^\dag}|0\rangle
\ee
So we have derived
\be\label{trace}
\Tr_a \hat{O} = \frac{1}{\pi}\int dq_1dq_2\left(e^{-|q|^2}\langle0|e^{\bar qa}\hat{O}e^{qa^\dag}|0\rangle\right)
\ee
Consider a matrix element of $D(k)$
\be\label{Gen}
G(\bar k, q; p) \equiv e^{\frac{1}{2}\left(|k|^2 + |q|^2\right)} \langle k | D(p)|q \rangle = e^{-\frac{|p|^2}{2}}e^{\bar k q + \bar k p - \bar q p}
\ee
Inserting resolution of unity in terms $|n\rangle$ we find
\be
G(\bar k, q; p) =\sum_{m,n} \bar{u}_m(k)u_n(q)D_{mn}(p)\,,
\ee
where
\be
u_n(k) \equiv \langle n|k\rangle=\frac{k^m}{\sqrt{m!}}
\ee
$G$ is a generating function of the matrix elements of $D_{mn}(p)$. The latter are obtain expanding (\ref{Gen}) in series in $k$ and $q$.
\bea
&&\!\!\!\!\!\!\!\!\!\!\!\!\!D_{nm}(p) = \sqrt{\frac{n!}{m!}} e^{-\frac{|p|^2}{2}}p^{m-n} L^{m-n}_n(|p|^2)\,,\,\,\, m\geq n \\
&&\!\!\!\!\!\!\!\!\!\!\!\!\!D_{nm}(p) = \sqrt{\frac{m!}{n!}} e^{-\frac{|p|^2}{2}}p^{n-m} L^{n-m}_m(|p|^2) \,,\,\,\, n\geq m
\eea
We also find simple relations for the traces 
\bea
&&\Tr D(p) = \pi \delta^{(2)}(p)\,\\
&&\Tr \left[D(p)D^{-1}(q)\right] = \pi \delta^{(2)}(p-q)\,.
 \eea

\subsection{Application: Trace over $b$-subspace}

We want to evaluate trace of a product of local operators
\be
\Tr_b \left[\prod_{i=1}^M O_i(x_i)\right] =\int [dk]\prod_{i=1}^M  \left[\Tr_b\left[e^{i\sum_{i}{\bf{k}}_i\cdot {\bf{x}}_i}\right]O_i(k_i)\right],
\ee
where we have introduce a shorthand notation $[dk] = \prod_{i=1}^M \frac{d^2{\bf{k}}_i}{(2\pi)^{2M}}$ and $O_i(k_i)$ is understood as a Fourier transform of $O_i(x_i)$. Now we re-write the exponent in terms of $a$ and $b$.
\be
e^{i{\bf{k}}\cdot {\bf{x}}} = e^{\frac{\bar k \ell}{\sqrt{2}}a-\frac{k\ell}{\sqrt{2}}a^\dag} e^{-\frac{\bar k \ell}{\sqrt{2}}b^\dag+\frac{k\ell}{\sqrt{2}}b} = e^{\bar q a-qa^\dag} e^{-\bar qb^\dag+qb}\,,
\ee
where we introduced $q = \frac{k\ell}{\sqrt{2}}$, so that $[dk] = \left(\frac{2}{\ell^2}\right)^M[dq]$. We have for the exponent
\be
e^{i{\bf{k}}\cdot {\bf{x}}} = D_a(-q)e^{-\frac{|q|^2}{2}}e^{-\bar qb^\dag}e^{ qb}\,.
\ee

Now, we plug this back into the trace
\be
\left(\frac{2}{l^2}\right)^M\prod_{i=1}^M\int [dq]  D_a(-q_i)\left[\Tr_b\left[e^{-\frac{|q_i|^2}{2}}e^{-\bar q_ib^\dag}e^{ q_ib}\right]O_i(q_i)\right]\,.
\ee

To proceed we use (\ref{trace})
\bea\nonumber
&&\Tr_b\left[\prod_{i=1}^Me^{-\frac{|q_i|^2}{2}}e^{-\bar q_ib^\dag}e^{ q_ib}\right]  \\\nonumber
&&=\frac{1}{\pi}e^{-\sum_i\frac{|q_i|^2}{2}}\int d^2p\left[e^{-|p|^2}\langle0|e^{\bar pb}\prod_{i=1}^Me^{-\bar q_ib^\dag}e^{ q_ib}e^{pb^\dag}|0\rangle\right]\,.
\eea

We want to normal order the product. In order to do this we use permutation relations
\bea\nonumber
&&e^{\bar pb}\prod_{i=1}^Me^{-\bar q_ib^\dag}e^{ q_ib}e^{pb^\dag} =\\\nonumber
 &&:e^{\bar pb}\prod_{i=1}^Me^{-\bar q_ib^\dag}e^{ q_ib}e^{pb^\dag}:e^{|p|^2}e^{\sum_{i>j}-\bar q_iq_j}e^{-\bar p\sum_i\bar q_i}e^{p\sum_iq_i}\,.
\eea

Denoting $\sum_i q_i = Q$ and using 
\be
\langle 0|:e^{\bar pb}\prod_{i=1}^Me^{-\bar q_ib^\dag}e^{ q_ib}e^{pb^\dag}:|0\rangle=1
\ee
 we have
\bea\nonumber
&&\Tr_b\left[\prod_{i=1}^Me^{-\frac{|q_i|^2}{2}}e^{-\bar q_ib^\dag}e^{ q_ib}\right] \\
&&= \frac{1}{\pi}e^{-\sum_i\frac{|q_i|^2}{2}}e^{\sum_{i>j}-\bar q_iq_j}\int d^2pe^{-\bar{p}Q}e^{p\bar Q}\,.
\eea

The latter integral is a $\delta$-function
\be
\frac{1}{\pi}\int dp_1dp_2e^{-\bar{p}Q}e^{p\bar Q}=\pi \delta^{(2)}({\bf{Q}})=\pi\int \frac{d^2{\bf{\lambda}}}{(2\pi)^2}e^{i{\bf{\lambda}}\cdot{\bf{Q}}}\,.
\ee

We also use $\bar q_i q_j = {\bf{q_i}} \cdot {\bf{q_j}} + i {\bf{q_i}} \wedge {\bf{q_j}}$, where ${\bf{a}}\wedge {\bf{b}} = a_1b_2 - a_2b_1$. As well as
\be
\frac{1}{2}\sum_i |q_i|^2 + \sum_{i<j} {\bf{q_i}}\cdot{\bf{q_j}} =\frac{1}{2} {\bf{Q}}^2
\ee
then
\be
\Tr_b\left[\prod_{i=1}^Me^{-\frac{|q_i|^2}{2}}e^{-\bar q_ib^\dag}e^{ q_ib}\right]  = \pi e^{-i\sum_{i>j}{\bf{q_i}}\wedge{\bf{q_j}}}\int \frac{d^2{\bf{\lambda}}}{(2\pi)^2}e^{i{\bf{\lambda}}\cdot{\bf{Q}}}
\ee

We have proven that
\begin{widetext}
\be
\Tr_b \left[\prod_{i=1}^M O_i(x_i)\right] =\pi \left(\frac{2}{l^2}\right)^M\int \frac{d^2{\bf{\lambda}}}{(2\pi)^2}\prod_{i=1}^M\int [dq]\left[  D_a(-q_i) e^{i{\bf{\lambda}}\cdot{\bf{q_i}}}O_i(q_i)\right]e^{-i\sum_{i>j}{\bf{q_i}}\wedge{\bf{q_j}}}\,.
\ee
Finally using
\be
D_a(-q_1) \cdot \ldots \cdot D_a(-q_M) = e^{i\sum_{i>j}q_i\wedge q_j}D_a(-q_1-\ldots-q_M)=e^{i\sum_{i>j}q_i\wedge q_j}D_a(-Q)
\ee
we arrive at the trace formula
\be
\Tr_b \left[\prod_{i=1}^M O_i(x_i)\right] =\pi \left(\frac{2}{l^2}\right)^M\int \frac{d^2{\bf{\lambda}}}{(2\pi)^2}\prod_{i=1}^M\int [dq]\left[ e^{i{\bf{\lambda}}\cdot{\bf{q_i}}}\tilde{O}_\ell i(q_i)\right]\,.
\ee
This is the generalization of the $b$-summation formula that was used in Appendix B to arbitrary number of external legs. This formula is useful if one is aiming to evaluate the generating functional to arbitrary order in external fields.
\end{widetext}
\section{Vertices for the Dirac polarization tensor}
In this Appendix we explicitly write out the vertices for the Dirac polarization tensor. These are obtained by straightforwardly combining \eqref{psinm} with \eqref{GammaD}, for  $|n|,|n'|>0$
\begin{widetext} 
\be
	\Gamma^0_{Dnn'}(\vec{k})=\frac{1}{2}\left(\sqrt{\frac{|n|!}{|n'|!}}\left(\frac{\bar{k}\ell }{\sqrt{2}}\right)^{|n'|-|n|}L_{|n|}^{|n'|-|n|}\left(\frac{|k\ell |^2}{2}\right)
	+\sign(n)\sign(n')\sqrt{\frac{(|n|-1)!}{(|n'|-1)!}}\left(\frac{\bar{k}\ell }{\sqrt{2}}\right)^{|n|'-|n|}L_{|n|-1}^{|n'|-|n|}\left(\frac{|k\ell |^2}{2}\right)\right)
\ee
\bea\nonumber
	\Gamma^1_{Dnn'}(\vec{k})=\frac{v_F}{2} \Big(\sign(n') && \sqrt{\frac{|n|!}{(|n'|-1)!}}\left(\frac{\bar{k}\ell }{\sqrt{2}}\right)^{|n'|-|n|-1}L_{|n|}^{|n'|-|n|-1}\left(\frac{|k\ell |^2}{2}\right)\\ 
	&&+\sign(n)\sqrt{\frac{(|n|-1)!}{(|n'|)!}}\left(\frac{\bar{k}\ell }{\sqrt{2}}\right)^{|n|'-|n|+1}L_{|n|-1}^{|n'|-|n|+1}\left(\frac{|k\ell |^2}{2}\right)\Big),
\eea
	\begin{align}
	\Gamma^2_{Dnn'}(\vec{k})=-\frac{iv_F}{2}&\left(\sign(n')\sqrt{\frac{|n|!}{(|n'|-1)!}}\left(\frac{\bar{k}\ell }{\sqrt{2}}\right)^{|n'|-|n|-1}L_{|n|}^{|n'|-|n|-1}\left(\frac{|k\ell |^2}{2}\right)\right.\nonumber\\
	&\left.-\sign(n)\sqrt{\frac{(|n|-1)!}{(|n'|)!}}\left(\frac{\bar{k}\ell }{\sqrt{2}}\right)^{|n|'-|n|+1}L_{|n|-1}^{|n'|-|n|+1}\left(\frac{|k\ell |^2}{2}\right)\right).
	\end{align}  
\end{widetext}
For the case $n=0$, $|n'|>0$
\begin{align}
\Gamma^0_{D0n'}(\vec{k})=\frac{1}{\sqrt{2}}\left(\sqrt{\frac{1}{|n'|!}}\left(\frac{\bar{k}\ell }{\sqrt{2}}\right)^{|n'|}L_{0}^{|n'|}\left(\frac{|k\ell |^2}{2}\right)\right),
\end{align}
\begin{align}
\Gamma^1_{D0n'}(\vec{k})=\frac{v_F}{\sqrt{2}}&\left(\sign(n')\sqrt{\frac{1}{(|n'|-1)!}}\left(\frac{\bar{k}\ell }{\sqrt{2}}\right)^{|n'|-1}\right.\nonumber\\&\times\left.L_{0}^{|n'|-1}\left(\frac{|k\ell |^2}{2}\right)\right),
\end{align}    
\begin{align}
\Gamma^2_{D0n'}(\vec{k})=-\frac{iv_F}{\sqrt{2}}&\left(\sign(n')\sqrt{\frac{1}{(|n'|-1)!}}\left(\frac{\bar{k}\ell }{\sqrt{2}}\right)^{|n'|-1}\right.\nonumber\\&\times\left.L_{0}^{|n'|-1}\left(\frac{|k\ell |^2}{2}\right)\right),
\end{align}
For the case $|n|>0$, $n'=0$
\begin{align}
\Gamma^0_{Dn0}(\vec{k})=\frac{1}{\sqrt{2}}\left(\sqrt{\frac{1}{|n|!}}\left(\frac{-k\ell }{\sqrt{2}}\right)^{|n|}L_{0}^{|n|}\left(\frac{|k\ell |^2}{2}\right)\right),
\end{align}
\begin{align}
\Gamma^1_{Dn0}(\vec{k})=\frac{v_F}{\sqrt{2}}&\left(\sign(n)\sqrt{\frac{1}{(|n|-1)!}}\left(\frac{-k\ell }{\sqrt{2}}\right)^{|n|-1}\right.\nonumber\\&\times\left.L_{0}^{|n|-1}\left(\frac{|k\ell |^2}{2}\right)\right),
\end{align}    
\begin{align}
\Gamma^2_{Dn0}(\vec{k})=\frac{iv_F}{\sqrt{2}}&\left(\sign(n)\sqrt{\frac{1}{(|n|-1)!}}\left(\frac{-k\ell }{\sqrt{2}}\right)^{|n|-1}\right.\nonumber\\&\times\left.L_{0}^{|n|-1}\left(\frac{|k\ell |^2}{2}\right)\right),
\end{align}
we can write down the explicit form of equation (\ref{eq:polDirac}) as the summation of product of Laguerre polynomials for each pair of indices $\mu$ and $\nu$. 

	\section{Evaluation of the infinite sums}
	\label{sec:DiracEx}
	In this Appendix, we again use $\vec{k}=(k_1,0)$ . The components of polarization tensors can be obtained from equation (\ref{eq:polDirac}) and the explicit form of vertex operator $\Gamma^\mu_{Dnn'}(\vec{k})$
	\begin{equation}
	\Pi^{12}_D(\Omega,\vec{k})=i\Omega \Pi^{12}_1+i\Omega k_1^2 \Pi^{12}_2+i\Omega^3 \Pi^{12}_3+\cdots,
	\end{equation} 
	where $\Pi^{12}_1$,$\Pi^{12}_2$ and $\Pi^{12}_3$ are the result of Taylor expansion of  (\ref{eq:polDirac}) at specific order of $\omega$ and $p$, $\cdots$ represents the higher order of frequency and momentum. The explicit form of $\Pi^{12}_1$ is 
	\begin{align}
	\Pi^{12}_1=&\frac{i}{8\pi }\left(\sum_{n=N+1}^{\infty} \left[(\sqrt{n}-\sqrt{n-1})^2-(\sqrt{n+1}-\sqrt{n})^2\right] \right.\nonumber\\
	&\left.+(\sqrt{N}+\sqrt{N+1})^2\right),
	\end{align}
	the first two terms come from the summation with $n'<0$ and $n>0$, the last term is from $n,n'>0$. Similarly,
	\begin{widetext}  
		\begin{align}
		\Pi^{12}_2=\frac{i\ell ^2}{64\pi }&\left(\sum_{n=N+1}^{\infty} \left[4(2n+1)(\sqrt{n+1}-\sqrt{n})^2-4(2n-1)(\sqrt{n}-\sqrt{n-1})^2\right.\right.\nonumber\\
		&\left.+(n-1)(\sqrt{n}-\sqrt{n-2})^2 -(n+1)(\sqrt{n+2}-\sqrt{n})^2\right]\nonumber\\ 
		&\left. -4(2N-1)(\sqrt{N+1}+\sqrt{N})^2+N(\sqrt{N+1}+\sqrt{N-1})^2+(N+1)(\sqrt{N+2}+\sqrt{N})^2\right)\,,
		\end{align}
	\end{widetext}
	\begin{align}
	\Pi^{12}_3=&\frac{i\ell ^2}{16\pi v_F^2}\left(\sum_{n=N+1}^{\infty} \left[(\sqrt{n}-\sqrt{n-1})^4-(\sqrt{n+1}-\sqrt{n})^4\right] \right.\nonumber\\
	&\left.+(\sqrt{N}+\sqrt{N+1})^4\right).
	\end{align}
	The summations are convergent and can evaluated $\Pi^{12}_1$, $\Pi^{12}_2$ and $\Pi^{12}_3$ to obtain 
	\begin{align}
	\Pi^{12}(\Omega,\vec{k})=&i\Omega \frac{N+1/2}{2\pi}-i\Omega k_1^2 \ell ^2 \frac{6N^2+6N+1}{16\pi}\nonumber\\
	&i\Omega^3 \frac{\ell ^2}{v_F^2}\frac{8N^2+8N+1}{8\pi}+\cdots.
	\end{align}
	We derive similarly
	\begin{align}
	\Pi^{00}_D(\Omega,\vec{k})=k_1^2\Pi^{00}_1+\cdots\,,
	\end{align}
	\begin{align}
	\Pi^{11}_D(\Omega,\vec{k})=\Omega^2\Pi^{11}_1+\cdots\,,
	\end{align}
	\begin{align}
	\Pi^{22}_D(\Omega,\vec{k})=\Omega^2\Pi^{22}_1+k_1^2\Pi^{22}_2+\cdots\,.
	\end{align}
	There is no $k_1^2$ term in $\Pi^{11}(\Omega,\vec{k})$ and no $\Omega^2$ term in $\Pi^{00}(\Omega,\vec{k})$, we can calculate the coefficients
	\begin{align}
	&\Pi^{00}_1=\Pi^{11}_1=\Pi^{22}_1=\nonumber\\
	&\frac{\ell }{8\sqrt{2}\pi v_F}\left(\sum_{n=N+1}^{\infty}\left[(\sqrt{n+1}-\sqrt{n})^3+(\sqrt{n}-\sqrt{n-1})^3\right]\right.\nonumber\\
	&\left.+(\sqrt{N}+\sqrt{N+1})^3\right)\,.\nonumber\\
	\end{align}
	The summation is convergent and is given by 
	\begin{align}
	\sum_{n=N+1}^{\infty}\left[(\sqrt{n+1}-\sqrt{n})^3+(\sqrt{n}-\sqrt{n-1})^3\right]=\nonumber\\-(\sqrt{N}+\sqrt{N+1})^3-12\zeta(-\frac{1}{2},N+1), 
	\end{align}
	where $\zeta(s,n)$ is the Hurwitz $\zeta$-function, which is defined as 
	\begin{eqnarray}
	\zeta(s,q)=\sum_{n=0}^{\infty}{\frac{1}{(n+q)^s}}.
	\end{eqnarray}
	As the result, we have 
	\begin{equation}
	\Pi^{00}_1=\Pi^{11}_1=\Pi^{22}_1=-\frac{3\ell }{2\sqrt{2}\pi v_F}\zeta(-\frac{1}{2},N+1)\,.
	\end{equation}

	The coefficient $\Pi^{22}_2$ can be also calculated similarly 
	\begin{equation}
	\Pi^{22}_2=\frac{3\ell v_F}{4\sqrt{2}\pi }\zeta(-\frac{1}{2},N+1).
	\end{equation}
We summarize these results in the Section \ref{sec:PolDirac}.

\section{Evaluate the summations of non-relativistic polarization tensor at large N limit}
In this Appendix, we will show the explicit calculation of polarization tensors in the Section \ref{sec:largeN}.
The following Bessel function identities will come in handy
\begin{equation}
\label{eq:bes0}
J_{1-\omega}(q)J_{1+\omega}(q)-J_{-1-\omega}(q)J_{-1+\omega}(q)=\frac{4\omega\sin(\pi\omega)}{\pi q^2}.
\end{equation}	
\begin{equation}
\label{eq:bes2}
\omega J_\omega(q)=\frac{q}{2}(J_{1+\omega}(q)+J_{-1+\omega}(q))
\end{equation}
\begin{equation}
\label{eq:bes3}
\omega J_{-\omega}(q)=-\frac{q}{2}(J_{1-\omega}(q)+J_{-1-\omega}(q))
\end{equation}
The summations (\ref{eq:xx1}),(\ref{eq:yy1}) and (\ref{eq:xy1}) can be recast as 
\label{sec:largeN-NRsum}
\begin{equation}
\Pi^{11}(q,\omega)=-\frac{N\omega_c}{2\pi}+\sum_{n=1}^{\infty}-\frac{2N n^4  \omega_c \left[J_n(q)\right]^2}{\pi q^2 (\omega^2-n^2)}.
\end{equation} 
\begin{align}
\label{eq:piyy}
\Pi^{22}(q,\omega)=&-\frac{N\omega_c}{2\pi}+\sum_{n=1}^{\infty}-\frac{N n^2 \omega_c \left[J_{n-1}(q)-J_{n+1}(q)\right]^2}{2\pi(\omega^2-n^2)}\nonumber\\
&-\rg\sum_{n=1}^{\infty}\frac{ n^2 \omega_c q \left[J_{n-1}(q)-J_{n+1}(q)\right]J_{n}(q)}{4\pi (\omega^2-n^2)}\nonumber\\
&-\rg^2\sum_{n=1}^{\infty}\frac{q^2  n^2 \omega_c \left[J_n(q)\right]^2}{32 N\pi (\omega^2-n^2)}\nonumber\\
=&-\frac{N\omega_c}{2\pi}-\frac{N\omega_c}{2\pi}\sum_{n=1}^{\infty}\frac{ n^2  \left[J_{n-1}(q)-J_{n+1}(q)\right]^2}{ (\omega^2-n^2)}\nonumber\\
&-\rg \omega_c q\frac{\partial}{\partial q}\sum_{n=1}^{\infty}\frac{ n^2  \left[J_{n}(q)\right]^2}{4\pi (\omega^2-n^2)}\nonumber\\
&-\rg^2\sum_{n=1}^{\infty}\frac{q^2  n^2 \omega_c \left[J_n(q)\right]^2}{32 N\pi (\omega^2-n^2)}\nonumber\\
=&-\frac{N\omega_c}{2\pi}-\frac{N\omega_c}{2\pi}\sum_{n=1}^{\infty}\left[\frac{ -4 n^2  J_{n-1}(q)J_{n+1}(q)}{ (\omega^2-n^2)}\nonumber\right. \\&\left.+\frac{ 4 n^4  J_{n}(q)J_{n}(q)}{ q^2(\omega^2-n^2)}\right]\nonumber\\
&-\rg \omega_c q\frac{\partial}{\partial q}\sum_{n=1}^{\infty}\frac{ n^2  \left[J_{n}(q)\right]^2}{4\pi (\omega^2-n^2)}\nonumber\\
&-\rg^2\sum_{n=1}^{\infty}\frac{q^2  n^2 \omega_c \left[J_n(q)\right]^2}{32 N\pi (\omega^2-n^2)}\nonumber\\
=&\Pi^{11}(q,\omega)+\frac{2N \omega_c}{\pi}\sum_{n=1}^{\infty}\frac{n^2  J_{n-1}(q)J_{n+1}(q)}{ (\omega^2-n^2)}\nonumber\\
&-\rg \omega_c q\frac{\partial}{\partial q}\sum_{n=1}^{\infty}\frac{ n^2  \left[J_{n}(q)\right]^2}{4\pi (\omega^2-n^2)}\nonumber\\
&-\rg^2\sum_{n=1}^{\infty}\frac{q^2  n^2 \omega_c \left[J_n(q)\right]^2}{32 N\pi (\omega^2-n^2)}\nonumber\\\,,
\end{align}
\begin{align}
\label{eq:pixy}
\Pi^{12}(q,\omega)=&\sum_{n=1}^{\infty}-\frac{i N n^2  \omega\omega_c J_n(q)\left[J_{n-1}(q)-J_{n+1}(q)\right]}{\pi q(\omega^2-n^2)}\nonumber\\
&-\rg\sum_{n=1}^{\infty}\frac{i\omega n^2 \omega_c \left[J_n(q)\right]^2}{4\pi (\omega^2-n^2)}\nonumber\\
=&-\frac{iN \omega\omega_c }{\pi q}\frac{\partial}{\partial q}\sum_{n=1}^{\infty}\frac{ n^2   J_n(q) J_n(q)}{(\omega^2-n^2)}\nonumber\\
&-\rg\sum_{n=1}^{\infty}\frac{i\omega n^2 \omega_c \left[J_n(q)\right]^2}{4\pi (\omega^2-n^2)}\nonumber\\
\end{align}
where we used 
\begin{equation}
\frac{\partial}{\partial x}J_n(x)=\frac{1}{2}(J_{n-1}(x)-J_{n+1}(x)).
\end{equation}	
Using the identity
\begin{equation}
\sum_{n=1}^{\infty}n^2 \left[J_n(x)\right]^2=\frac{x^2}{4}
\end{equation} 
we can rewrite $\Pi^{11}(q,\omega)$ as 
\begin{equation}
\label{eq:pixx}
\Pi^{11}(q,\omega)=-\frac{2N \omega^2  \omega_c}{\pi q^2}\sum_{n=1}^{\infty}\frac{ n^2 \left[J_n(q)\right]^2}{ (\omega^2-n^2)}.
\end{equation}
Next we need to evaluate 
\begin{align}
\sum_{n=1}^{\infty}\frac{ n^2 \left[J_n(q)\right]^2}{ (\omega^2-n^2)}=\omega^2\sum_{n=1}^{\infty}\frac{  \left[J_n(q)\right]^2}{ (\omega^2-n^2)}-\sum_{n=1}^{\infty}  \left[J_n(q)\right]^2,
\end{align}
\begin{align}
\sum_{n=1}^{\infty}\frac{ n^2 J_{n-1}(q)J_{n+1}(q)}{ (\omega^2-n^2)}=&\omega^2\sum_{n=1}^{\infty}\frac{  J_{n-1}(q)J_{n+1}(q)}{ (\omega^2-n^2)}\nonumber\\&-\sum_{n=1}^{\infty}J_{n-1}(q)J_{n+1}(q)
\end{align}	
to derive the closed form of $\Pi^{ij}(p,\omega)$. Both of the above summations can be evaluated using the tricks in reference \cite{Simon1993}, which gives us
\begin{equation}
\sum_{n=1}^{\infty}\frac{ n^2 \left[J_n(q)\right]^2}{ (\omega^2-n^2)}=-\frac{1}{2}+\frac{\pi \omega}{2\sin(\pi\omega)}J_{\omega}(q)J_{-\omega}(q)
\end{equation}
\begin{equation}
\sum_{n=1}^{\infty}\frac{ n^2 J_{n-1}(q)J_{n+1}(q)}{ (\omega^2-n^2)}=-\frac{\pi \omega}{2\sin(\pi\omega)}J_{1-\omega}(q)J_{1+\omega}(q)
\end{equation}
We therefore can derive the closed form of polarization tensor
\bea
\label{eq:LargeNxx}
\!\Pi^{11}(q,\omega)&=&\frac{N \omega^2  \omega_c}{\pi q^2}\left(1-\frac{\pi \omega}{\sin(\pi\omega)}J_{\omega}(q)J_{-\omega}(q)\right),\nonumber\\\\
\label{eq:LargeNxy}
\Pi^{12}(q,\omega)&=&-\frac{i N \omega\omega_c }{2\pi q}\frac{\pi \omega}{\sin(\pi\omega)}\frac{\partial}{\partial q}\left[J_{\omega}(q)J_{-\omega}(q)\right]\nonumber\\&&+i\frac{g  \omega \omega_c}{8 \pi } \left(1-\frac{\pi \omega}{\sin(\pi\omega)}J_{\omega}(q)J_{-\omega}(q)\right),\nonumber\\\\
\label{eq:LargeNyy}
\Pi^{22}(q,\omega)&=&\frac{N \omega^2  \omega_c}{\pi q^2}\left(1-\frac{\pi \omega}{\sin(\pi\omega)}J_{\omega}(q)J_{-\omega}(q)\right)\nonumber\\&&-\frac{N \omega_c \omega}{\sin(\pi\omega)}J_{1-\omega}(q)J_{1+\omega}(q)\nonumber\\&&-\frac{g q \omega  \omega c}{8 \sin (\pi  \omega )}\frac{\partial}{\partial q}\left[J_{\omega}(q)J_{-\omega}(q)\right]+\nonumber\\
&&+\frac{g^2 q^2 \text{$\omega $c}}{64 \pi  \text{N}}\left(1-\frac{\pi \omega}{\sin(\pi\omega)}J_{\omega}(q)J_{-\omega}(q)\right).\nonumber\\
\eea
Since the closed form of polarization tensor is obtained, we can derive the large $N$ approximation of conductivity and compare with Fermi liquid calculation.  

	\bibliography{landaupolar-v1}{}

\end{document}